\documentclass[ reprint,
groupedaddress,
amsmath,amssymb,
aps,
pra,
onecolumn]{revtex4-2}

\usepackage{graphicx}
\usepackage{comment}
\usepackage{epsfig}
\usepackage{epstopdf}
\usepackage{subcaption}
\usepackage{color}
\usepackage{gensymb}
\usepackage[font=footnotesize,labelformat=simple]{subcaption}

\usepackage{dcolumn}
\usepackage{bm}
\usepackage{amsmath,amssymb}
\usepackage[colorinlistoftodos]{todonotes}

\def\.{\cdot}

\def\_#1{{\bf #1\mit}}
\def\=#1{\overline{\overline #1}}

\newcommand{\blue}[1]{\textcolor{blue}{#1}}

\graphicspath{{figures/}}
\begin{document}

\title{Photonic Temporal Illusion} 

\author{G.~Ptitcyn$^1$}
\email{grigoriiptitcyn@gmail.com}  
\author{D.~M.~Sol{\'i}s$^2$}
\email{dmartinezsolis@com.uvigo.es}
\author{M.~S.~Mirmoosa$^3$}
\email{m.s.mirmoosa@outlook.com}
\author{N.~Engheta$^1$}
\email{engheta@seas.upenn.edu}

\affiliation{$^1$Department of Electrical and Systems Engineering, University of Pennsylvania, Philadelphia, PA 19104, USA\\
$^2$Departamento de Teoría de la Se\~{n}al y Comunicaciones, University of Vigo, 36301 Vigo, Spain \\
$^3$Center for Photonics Sciences, University of Eastern Finland, P.O.~Box 111, FI-80101 Joensuu, Finland} 


\begin{abstract} 
Materials with unusual optical properties are central to advanced control of light. Yet, in nature, such materials may be exceedingly rare and often difficult to obtain. To overcome this limitation, here we introduce the concept of temporal illusion: A temporally dynamic framework in which carefully programmed temporal variations in effective parameters generate responses akin to those of, in principle, any arbitrary time-invariant structure. We theoretically demonstrate that proper modulation of the permittivity of a conventional dielectric in space and time replicates the optical behavior associated with exotic materials. Besides, we reveal that, beyond steady-state effects, temporal illusion also enables control over transient responses, for instance, by effectively lowering the time constant of high-quality-factor resonators, therefore, allowing faster energy accumulation. Moreover, by incorporating detuning between modulation and excitation, we show that the framework unlocks additional functionalities. The temporal illusion paradigm thus broadens the capabilities of space-time varying systems, offering a powerful route to synthesize material responses on demand and paving the way for new theoretical and experimental directions in optics and wave physics. 
\end{abstract}
\maketitle


\section{Introduction}  

Recently, time-varying systems have captured increasing attention in optics (e.g.,~Refs.~\cite{galiffi_photonics_2022,engheta2023four,caloz2019spacetimeI1,caloz2019spacetimeII2, Alu_TL,lustig2023timeEx,Shalaev2025spatio}), fueled by both conceptual breakthroughs and experimental advances~\cite{Alu_TL,lustig2023timeEx,Shalaev2025spatio}. From a theoretical perspective, deliberately embedding temporal variations (adiabatic or ultrafast) into a material, modeled with macroscopic parameters, which in turn may also exhibit spatial nonuniformity, introduces a new degree of freedom~\cite{engheta2020metamaterials}. Exploiting this freedom can reshape the interaction of light with matter across both classical~\cite{morgenthaler1958velocity,mendoncca2002time,Agrawal2014RTC} and quantum regimes~\cite{mendoncca2000quantum,mendoncca2003temporal,vazquez2022shaping,liberal2023quantum,mirmoosa2023quantum}, enabling unprecedented strategies for controlling wave dynamics and engineering quantum states of light. A rapidly expanding body of research on diverse optical platforms---including temporally local and nonlocal isotropic, anisotropic, and bianisotropic media~\cite{Tretyakov2024Bian} as well as metasurfaces---has revealed a spectrum of phenomena~\cite{zhou2020broadband,pacheco2020temporal,quinones_tunable_2021,yin2022efficient,biancalana_dynamics_2007,zurita-sanchez_reflection_2009,reyes-ayona_observation_2015,lustig_topological_2018,park_spatiotemporal_2021,sharabi2021disordered,GregAtom}, ranging from artificial magnetic field for photons~\cite{fang_realizing_2012} and optically induced negative refraction~\cite{vezzoli_optical_2018} to frequency conversion~\cite{salary2018time}, amplification~\cite{wang2018photonic,koutserimpas2018nonreciprocal}, Doppler shift~\cite{ramaccia2017doppler,ramaccia2019phase}, Fresnel drag~\cite{huidobro2019fresnel}, camouflage~\cite{liu2019time,wang2020spread}, and nonreciprocity~\cite{yu2009complete,sounas2014angular,shi2017optical,dinc2017synchronized,fleury2018non,Our2}, among many others.

Given this capacity of temporal modulation to tailor light-matter interactions, an intriguing question may arise, which, to the best of our knowledge, has not yet been posed. That is, ``\textit{Can temporal modulation offer a pathway to overcome one of the challenges in light-matter interaction, namely the absence of desired time-invariant materials with extreme or highly unconventional parameters (for instance, materials with exceedingly large refractive indices for high frequencies})?" To explore this possibility, it is informative to draw an analogy with circuit theory, where the study of temporally varying reactive elements has also been a central theme of research for decades (see, e.g.,~Refs.~\cite{cullen1958travelling,magierowski2010rf,gray2010analytical,Afshari2010noise,qin2014nonreciprocal,mirmoosa2019time,hedayati2021parametric}). Notably, it has been shown that a capacitance oscillating at twice the excitation frequency can effectively mimic the static equivalent of a capacitance in parallel with a positive-valued or negative-valued resistance (which means that the system exhibits dissipative or amplifying characteristics depending on the sign of resistance)~\cite[Ch.~10, p.~410]{PABook}. Building on this idea, it has been demonstrated more recently that, by properly designing the temporal modulation profile, time-varying reactive elements can replicate a wide spectrum of effective responses--including dissipative, amplifying, reactive, and even non-Foster behaviors~\cite{ptitcyn2023time}. Inspired by these insights from circuit theory, one gravitates to a compelling possibility in optics: Modulating effective material parameters in time may result in emulating the properties of time-invariant optical materials, particularly those that are otherwise challenging to find in nature.

In this paper, we establish that a medium with electric response, parameterized by a permittivity that changes simultaneously both in space and time, following a specific profile, can exhibit a scattering behavior analogous to that of any targeted (arbitrary) linear time-invariant medium. We refer to this effect of reproducing responses under spatiotemporal modulation as a {\it{photonic temporal illusion}}. To highlight its potential and implications, we apply our theoretical framework to three classes of materials  in one-dimensional wave propagation: (i) High-index materials, (ii) materials with purely imaginary index, and (iii) near-zero-index (NZI) materials. Accordingly, we discuss the underlying mechanisms of temporal illusion and their salient features, and also show that, due to the inherently active nature of these systems, the effect is highly tunable, enabling dynamic control and enhanced functionality. 

The paper is organized as follows: First, we present our theoretical framework. Then we apply this framework to analyze the three representative examples mentioned above. Finally, we summarize the main conclusions and discuss possible directions for future research. Throughout the paper, time-harmonic oscillations are represented by $\exp(j\omega t)$.

\section{{Theoretical Framework}}

\begin{figure}
\centering
\includegraphics[width=1\textwidth]{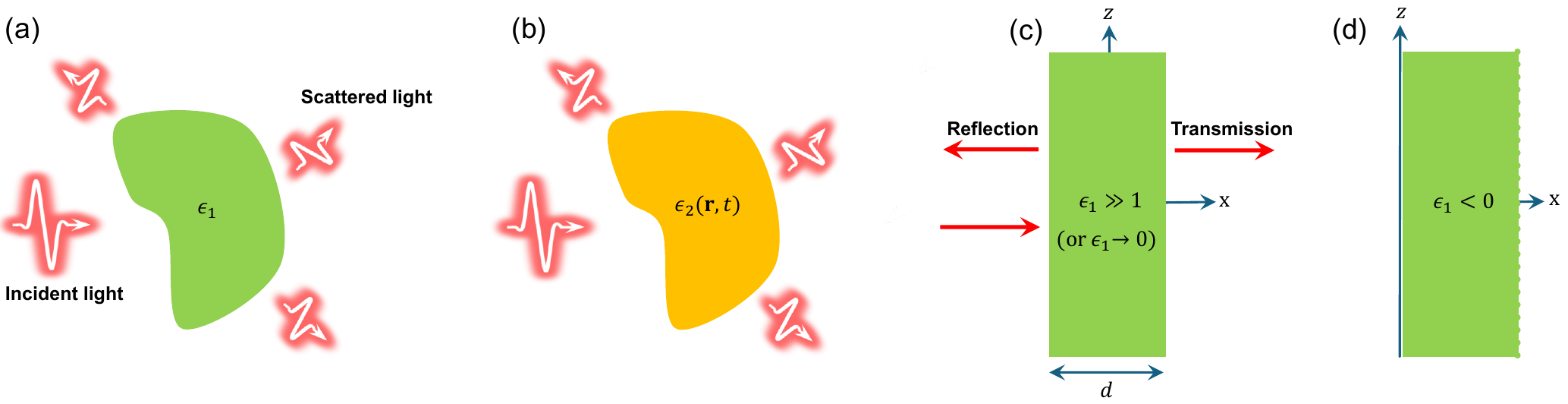}
\caption{\textbf{The concept of photonic temporal illusion.} (a) The linear and time-invariant object with permittivity $\epsilon_1$, which is under illumination of an electromanetic wave. (b) The electromagnetic response in (a) is reproduced by an object whose permittivity is modulated properly in space and time. To avoid singularity for the required relative permittivity, a DC electric field needs to be added. (c) and (d) Schematic views of the one-dimensional scenario for the three examples studied in Section~\ref{section:Results}.}
\label{fig:Fig1_conceptual}
\end{figure}

\noindent The realization of photonic temporal illusion can be rigorously established within the framework of Maxwell’s equations. According to the Maxwell-Amp{\`e}re law, in a source-free region, the curl of the magnetic field corresponds to the time derivative of the electric flux density: $\nabla\times\_H(\_r,t)=\partial_t\_D(\_r,t)$. By ignoring temporal nonlocality (i.e.,~assuming no dispersion) and considering only isotropic linear electric response, the electric flux density is given by the constitutive relation as $\_D(\_r,t)=\epsilon_0\epsilon(\_r,t)\_E(\_r,t)$, where $\epsilon_0$ and $\epsilon$ denote free-space permittivity and relative permittivity, respectively, and $\_E$ represents the electric field. This relation can be directly substituted into the Maxwell-Amp{\`e}re law. In stationary (i.e.,~time-invariant) media [i.e.,~$\epsilon(\_r,t)\equiv\epsilon(\_r)$, see Fig.~\ref{fig:Fig1_conceptual}(a)], permittivity is factored out of the time derivative, and the displacement current reduces to the familiar form $\partial_t\_D(\_r,t) = \epsilon_0\epsilon(\_r)\partial_t\_E(\_r,t)$.
In contrast, for nonstationary (i.e.,~time-variant) media [i.e., $\epsilon(\_r,t)$ varies explicitly with time, see Fig.~\ref{fig:Fig1_conceptual}(b)=], the time derivative of the permittivity itself contributes with an extra term as $\partial_t\_D(\_r,t) =\epsilon_0\Big[\epsilon(\_r,t)\partial_t\_E(\_r,t) +  \_E(\_r,t)\partial_t\epsilon(\_r,t)\Big]$. This distinction leads to the existence of two strikingly different media, which, however, can produce identical electric and magnetic fields. Indeed, if we label the stationary medium with subscript ``1" and the nonstationary medium with ``2", and require $\_E_1=\_E_2$, the corresponding displacement currents must satisfy $\partial_t\_D_1 = \partial_t\_D_2$, even though the flux densities themselves are allowed to differ, $\_D_1\neq\_D_2$.

Consequently, for light propagating through a time-invariant dielectric medium with relative permittivity $\epsilon_1$, a space-time-varying diagonal relative permittivity tensor $\=\epsilon_2(\_r,t)$ can, in principle, be constructed that reproduces identical electromagnetic fields $\_E_1(\_r,t)$ and $\_H_1(\_r,t)$ as those expected in the time-invariant medium with $\epsilon_1$, throughout space and time. After performing straightforward algebraic manipulation, the desired spatiotemporal relative permittivity profile is obtained as tensor components
\begin{align} 
\epsilon_{2,ll}(\_r,t)=\epsilon_1+\frac{Q_{l}(\_r)}{E_{1,l}(\_r,t)},
\label{EQ_space_time_permittivity} 
\end{align}
where the subscript $l$ denotes the respective vector component in the chosen basis.
In Eq.~\eqref{EQ_space_time_permittivity}, $\_Q$ is a static vector with dimensions of volt per meter ($\frac{{\rm{V}}}{{\rm{m}}}$), which can vary in space ``arbitrarily" (we will comment on this later). 
The space-time-varying term of the permittivity induces an additional polarization vector with $l$ component $\epsilon_0 Q_l \frac{E_{2,l}}{E_{1,l}}$. In general, this extra polarization is thus time variant too, since it follows the ratio $\frac{E_{2,l}}{E_{1,l}}$ by a time-invariant factor $\epsilon_0 Q_l$. However, when $\_E_1=\_E_2$, this factor becomes the true additional polarization, contributes to the actual displacement field as $\_D_2(\_r,t)=\epsilon_0\epsilon_1(\_r)\_E_1(\_r,t)+\epsilon_0\_Q(\_r)=\_D_1(\_r,t)+\epsilon_0\_Q(\_r)$, and yet, as initially required, has no effect on the displacement current. This is the core of our proposed temporal illusion concept. 

Three critical aspects of Eq.~\eqref{EQ_space_time_permittivity} need to be discussed in detail. First, the presence of the electric field components in the denominator dictates that if these cross zero, $\epsilon_2(\_r,t)$ will diverge to infinity, rendering practical implementation essentially impossible. This difficulty can be mitigated by introducing a temporally constant offset to the electric field, $\_E_{DC}$, with each $l$ component's magnitude exceeding the corresponding AC peak amplitude, thus ensuring that the denominator in Eq.~\eqref{EQ_space_time_permittivity} is always nonzero. This requirement can be regarded as the necessary cost---energy-wise, e.g.---of achieving the desired effect. (It is worth noting that, later in this work, due to the specific geometry of the problem under study and the resulting simplicity, this DC electric field is assumed to be uniform. However, in general, considering more complex geometries and problems, it can be spatially nonuniform while temporally static, therefore requiring zero curl: $\nabla \times \_E_{DC}(\_r)=\_0$.  Notice that, in contrast, $\_Q(\_r)$ does not need to be curl-free. Evidently, as prescribed by the boundary conditions---including continuity of normal \_D---, some form of electrostatic field distribution will be needed in the time-invariant surrounding as well, so that the temporal illusion phenomenon takes place.) Second, it is generally more feasible for the relative permittivity $\epsilon_2(\_r,t)$ to be greater than unity. This condition can be fulfilled by a judicious  choice of the ratios $Q/(E_{DC}\pm E_{AC})$ \textit{w.r.t.} $\epsilon_1$ (in which $E_{AC}$ refers to the AC amplitude). 

Third, a static medium characterized by $\epsilon_1$ and a space-time-varying medium with relative permittivity $\epsilon_2(\_r,t)$ result in the same responses ``at all times", provided that the modulation function identically follows the instantaneous electric field (in other words, $\_E_1=\_E_2\equiv\_E$). This condition implies that the modulating system must incorporate a dedicated feedback mechanism that continuously probes the field and simultaneously generates the corresponding modulation $\epsilon_2(\_r,t)$. For this reason, we designate this class of modulation as the ``instantaneous feedback" scheme. While it may nowadays be impractical with the current technology, such a mechanism is particularly ideal, as it in principle offers a perfect illusion of the medium with $\epsilon_1$. However, under continuous-wave excitations (e.g.,~time-harmonic sources), the transient response is typically irrelevant, and only the long-term equilibrium behavior is of interest. Thus, from this perspective, instead of relying on instantaneous feedback, one may construct $\epsilon_2(\_r,t)$ in Eq.~\eqref{EQ_space_time_permittivity} on the basis of the steady-state (equilibrium) solution of the desired field (which coincides with the instantaneous solution after a finite settling interval when the transient oscillations die out). In this work, we refer to this alternative strategy as the ``steady-state" modulation approach. Unlike the instantaneous feedback scheme, the steady-state approach does not involve any active feedback; rather, the modulating system must be pre-programmed with information about the target field distribution. It should also be noted that, from the perspective of an external observer, during the initial transient phase of the steady-state modulation, the illusion is not perfect until the steady state is reached. 

In the following, we will demonstrate the full potential of the photonic temporal illusion using three examples: (i) a dielectric slab with $\epsilon_1\gg1$, (ii) a half-space with $\epsilon_1<0$, and (iii) a slab of epsilon-near-zero medium with $0<\epsilon_1<1$, for the one-dimensional wave propagation.  The corresponding structures related to these examples have been shown in Figs.~1(c) and (d).

\begin{figure*}
\centering
\begin{subfigure}{0.24\textwidth}
\centering
 \includegraphics[width=\textwidth]{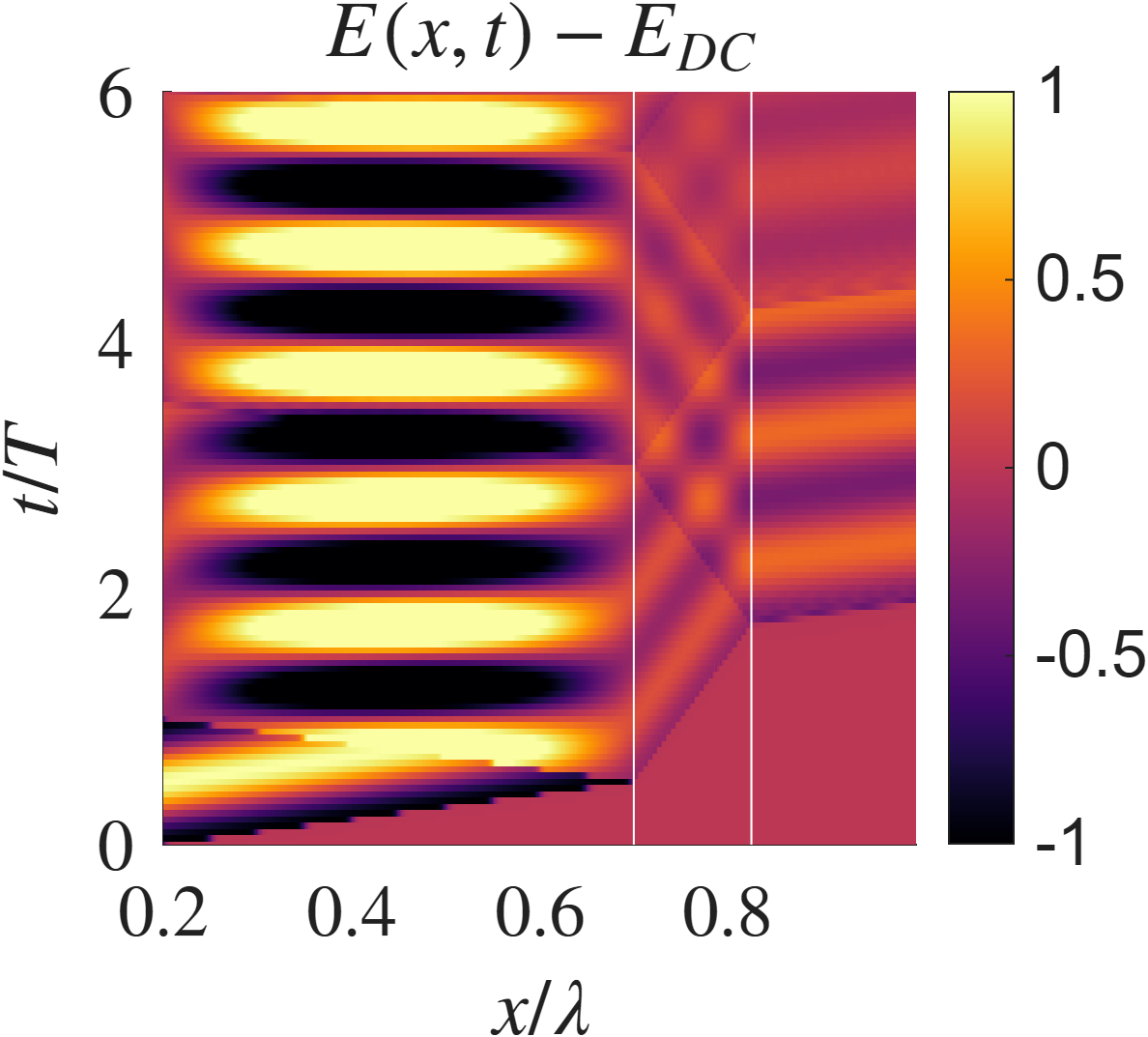}
\caption{}
\end{subfigure}
\begin{subfigure}{0.24\textwidth}
\centering
 \includegraphics[width=\textwidth]{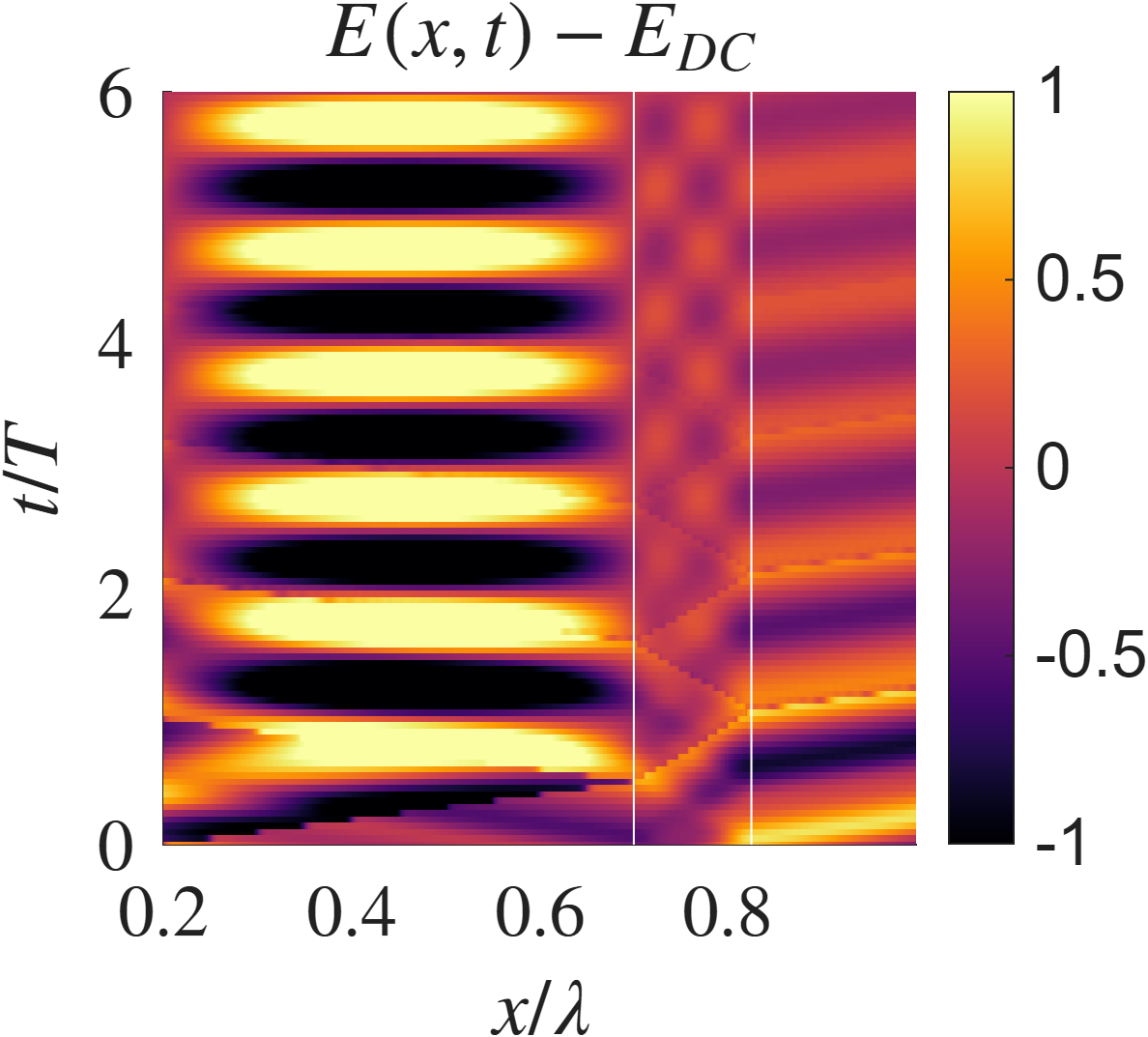}
\caption{}
\end{subfigure}
\begin{subfigure}{0.22\textwidth}
\centering
 \includegraphics[width=\textwidth]{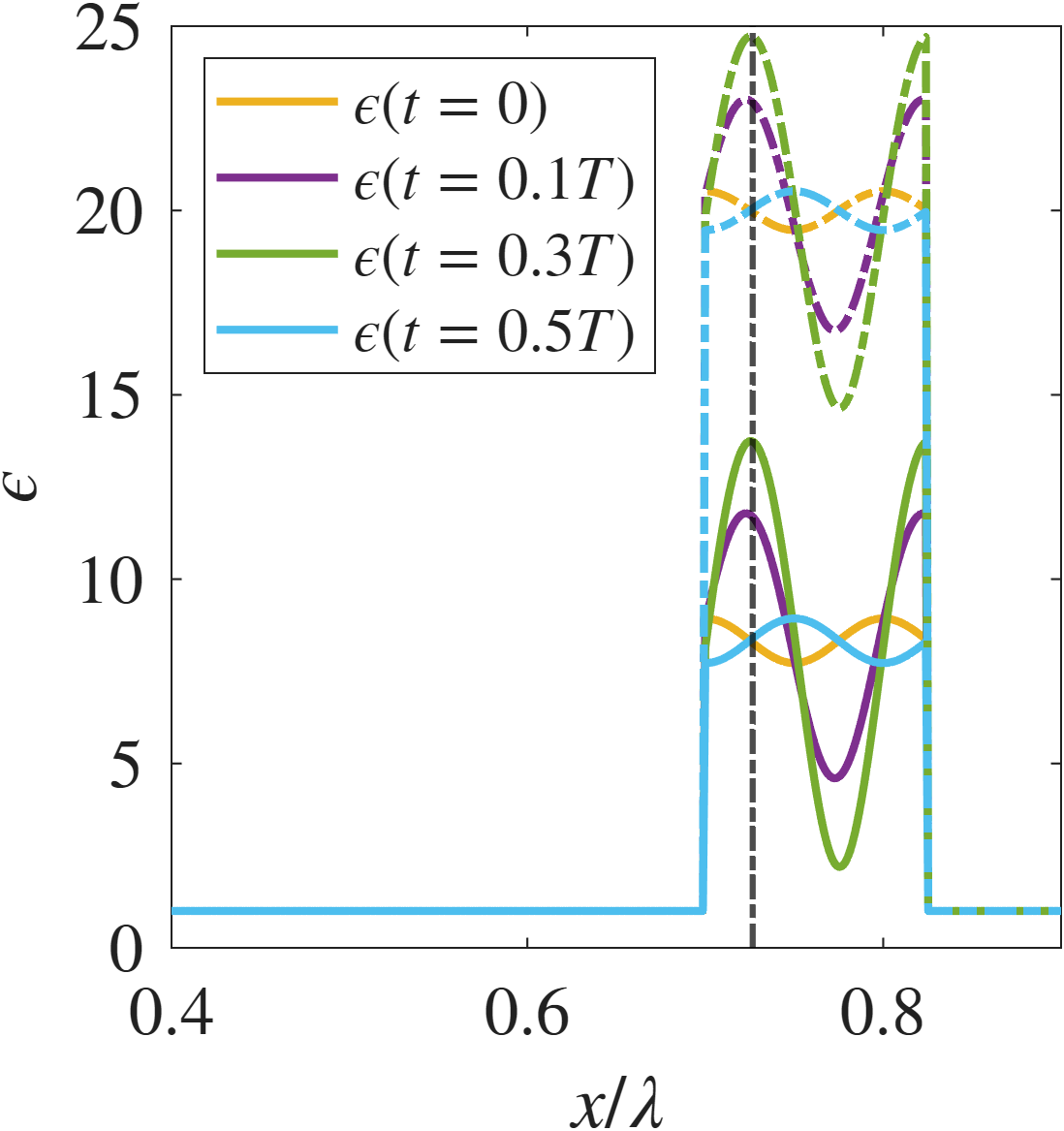}
\caption{}
\end{subfigure}
\begin{subfigure}{0.22\textwidth}
\centering
 \includegraphics[width=\textwidth]{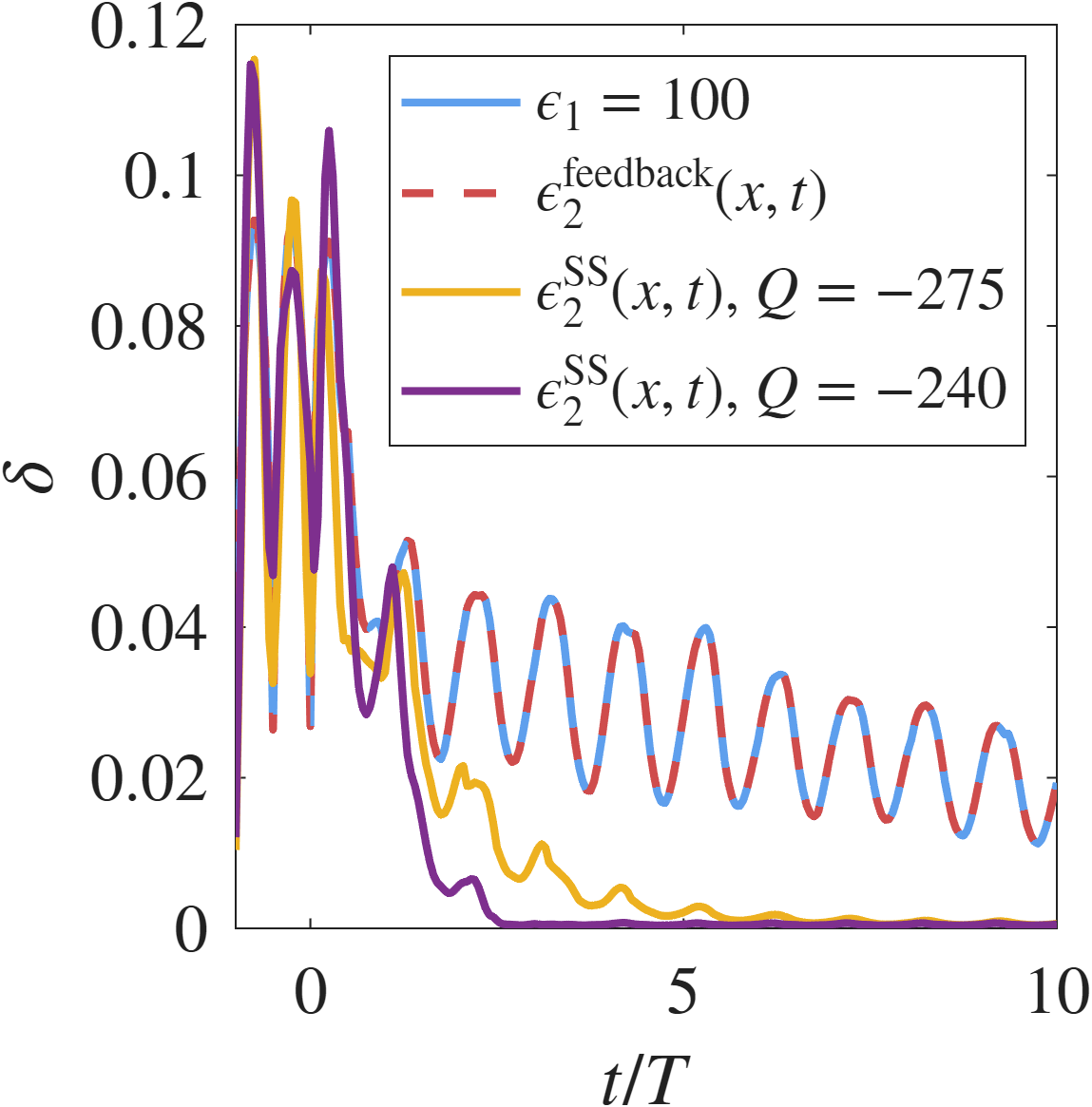}
\caption{}
\end{subfigure}
\caption{\textbf{Temporal illusion for a one-dimensional slab with an arbitrarily high $\epsilon_1$.} 
(a) Electric field in space and time shifted by $E_{DC}=3\frac{{\rm{V}}}{{\rm{m}}}$ for a slab with constant permittivity $\epsilon_1=100$ and thickness $d=\lambda/8$. (This panel relates to both cases of the exact theoretical result and the instantaneous feedback modulation.)  (b) Electric field in space and time shifted by $E_{DC}=3\frac{{\rm{V}}}{{\rm{m}}}$ for the same geometry as in (a), but for the steady-state modulation with dielectric permittivity varying in space and time according to $\epsilon_2(x,t)$ in steady state, so that it creates an illusion of $\epsilon_1=100$ after the transient response is passed ($Q=-240\frac{{\rm{V}}}{{\rm{m}}}$). 
(c) Spatiotemporal variation of permittivity of the slab $\epsilon_2(x,t)$ for constant $Q=-275\frac{{\rm{V}}}{{\rm{m}}}$ and $Q=-240\frac{{\rm{V}}}{{\rm{m}}}$. 
(d) Comparison among the rates of the convergence to the steady state  for four cases: constant $\epsilon_1=100$, space-time varying $\epsilon_2(x,t)$ in instantaneous feedback modulation scheme, and steady-state modulated $\epsilon_2(x,t)$ with different constants $Q$.} 

\label{fig_temporal_illusion}
\end{figure*} 


\section{Results} 
\label{section:Results}

\subsection{Illusion of arbitrarily high-index material with $\epsilon_1\gg1$}

\noindent At high frequencies, materials with high values of dielectric permittivity may not be readily available, nonetheless they are desirable. In this subsection, it is shown how to create the illusion of a dielectric slab with arbitrarily high relative permittivity $\epsilon_1\gg1$. For simplicity, in the rest of this paper we restrict our analysis to the one-dimensional (1D) uniform plane wave with frequency $\omega_0$, propagating along the x axis, and normally incident at the slab (for our numerical simulations we arbitrarily choose $\lambda=1$m).

As discussed earlier, two types of modulation mechanisms can be employed: Instantaneous feedback and steady-state, depending on the flexibility of the available space-time varying medium. We next highlight the differences between these two approaches and discuss the distinct advantages offered by each. Figure~\ref{fig_temporal_illusion}(a) presents the electric field in space and time for the time-invariant slab case. The instantaneous modulation scheme ensures matching of the fields with the time-invariant case at all moments of time. Therefore, the fields in Fig.~\ref{fig_temporal_illusion}(a) are identical to the fields obtained in the instantaneous modulation scenario. Figure~\ref{fig_temporal_illusion}(b) indicates the electric field in the case of a slab with space-time varying permittivity subject to a steady-state modulation scheme (see Supplementary information for derivations). Figure~\ref{fig_temporal_illusion}(c) shows the spatiotemporal modulation of permittivity $\epsilon_2(x,t)$ for two representative choices of the constant parameter, $Q=-240\frac{{\rm{V}}}{{\rm{m}}}$ and $Q=-275\frac{{\rm{V}}}{{\rm{m}}}$. Finally, Fig.~\ref{fig_temporal_illusion}(d) illustrates the difference as a function of time between the fields obtained analytically for the steady state and the fields obtained in simulations. To quantitatively assess this difference in a mathematically consistent fashion, we use the Euclidean norm to define the following time-dependent relative-error metric 
\begin{align}
\delta (t) = \frac{ \sqrt{ \int_0^L \Big[E(x,t) - E_{\rm{SS}}(x,t) \Big]^2 \mathrm{d}x } } { \sqrt{ \int_0^L E_{\rm{SS}}^2(x,t) \mathrm{d}x } },
\end{align}
with $L$ the length of the simulation domain ($L=1.15\lambda$ in this particular case), and where $E(x,t)$ denotes the numerically computed electric field and $E_{\rm{SS}}(x,t)$ represents the steady-state analytical solution. It is worth mentioning that integration is performed over the whole simulation domain, including regions inside and outside of the slab. 
Figure~\ref{fig_temporal_illusion}(d) reveals two key aspects of different modulation schemes: (1)~Error curves are identical for the time-invariant slab case (with $\epsilon_1 =100$) and instantaneous feedback modulation scheme, which means, as mentioned previously, that the fields in these cases are identical for all moments of time; (2)~Error converges to zero sooner for the steady-state modulation scheme (see transmission in Fig.~\ref{fig_temporal_illusion}(b)). For the time-invariant slab and instantaneous feedback cases, convergence is defined by $\epsilon_1$, which determines the quality factor of the slab.  For the steady-state modulation, convergence to zero depends on the spatiotemporal average of the modulation function $\epsilon_2(\_r,t)$, which can be engineered by choosing different $Q(\_r)$ (see Fig.~\ref{fig_temporal_illusion}(c)).
In Fig.~\ref{fig_temporal_illusion}(c), $Q(\_r)=-275\frac{{\rm{V}}}{{\rm{m}}}$ gives $\epsilon_2(\_r,t)$ as small as possible while staying larger than unity, whereas spatiotemporal average of $\epsilon_2(\_r,t)$ for $Q(\_r)=-240\frac{{\rm{V}}}{{\rm{m}}}$ is larger, which results in a longer convergence time. This reveals an interesting benefit of an illusion of arbitrarily high positive permittivity:  one can engineer the rate of convergence to the steady state, which has a potential in high-quality-factor resonators. Using conventional methods, one cannot pump huge energy into a resonator ``quickly", since time to the steady state depends on the quality factor. The temporal illusion may provide a pathway to address this issue. 


\subsubsection*{Homogeneous solutions in a DC-biased space-time modulated bulk}

\begin{figure}[h]
  \centering
  \includegraphics[width=16cm]{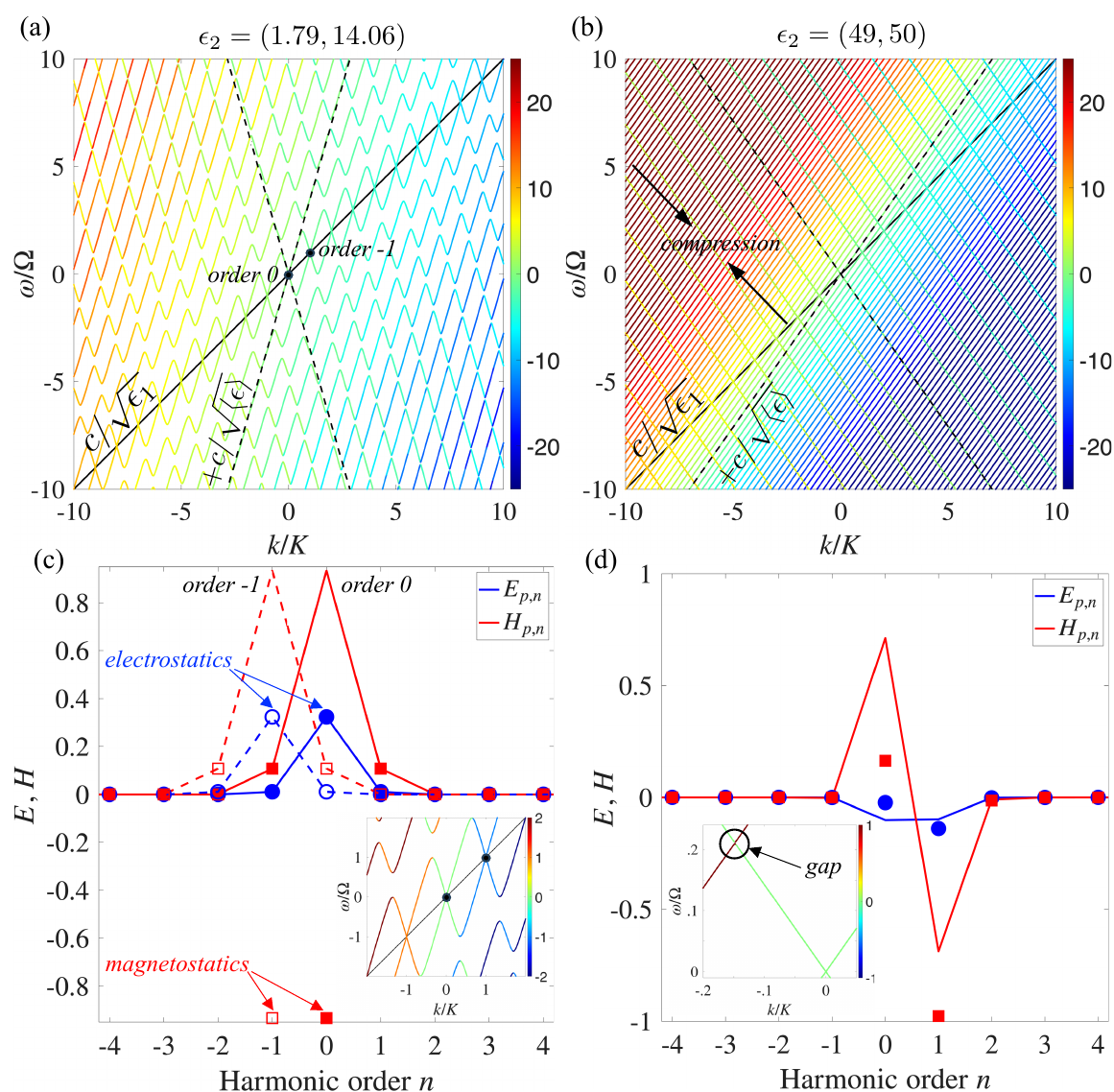}
\caption{\textbf{Bandgap structure of the proposed spacetime-modulated material.} (a),(b) Dispersion bands for two different sets of parameters: $Q=-275\frac{{\rm{V}}}{{\rm{m}}}$, $E_{AC}=0.2\frac{{\rm{V}}}{{\rm{m}}}$, and $Q=-151.4851\frac{{\rm{V}}}{{\rm{m}}}$, $E_{AC}=0.0297\frac{{\rm{V}}}{{\rm{m}}}$. In both cases, $\epsilon_1=100$ and $E_{DC}=3\frac{{\rm{V}}}{{\rm{m}}}$. The colorbar represents the dominant harmonic order of the associated eigenvector. (c),(d) Coefficients (purely real) of the Bloch expansion of the electric and magnetic eigenfields corresponding to coalescing eigenmodes in (a) and gap modes in (b), respectively.}\label{Fig_spacetime_dispersion}
\end{figure}

In order to gain further insights into our proposed modulation mechanism, particularly the steady-state approach, let us assume our dielectric function to be
\begin{equation}
\epsilon_2(x,t)=\epsilon_1+\frac{Q}{E_{DC} + E_{AC} \cos \left(\Omega t-Kx \right)}, 
\label{eq_epsilon_xt}
\end{equation}
with $\Omega=kc/\sqrt{\epsilon_1}$, i.e., the medium is modulated by a plane wave traveling along positive $x$ direction at the speed of light in the medium we are trying to emulate. Clearly, the Fourier spectrum of the space-time varying term of such function will be increasingly spread with the modulation strength $E_{AC}/E_{DC}$, whereas the ratio $Q/E_{DC}$, relative to $\epsilon_1-1$, roughly tells us the fraction of spacetime-modulation-induced polarization density. From $x$ and $t$ translation invariance, the transformed function can in general be expanded as a discrete double sum
\begin{equation}
\tilde{\epsilon}_2(k,\omega) = \sum_{m=-\infty}^{\infty} \sum_{n=-\infty}^{\infty} \tilde{\epsilon}_{m,n}\delta(\omega-m\Omega) \delta(k-nK),
\label{eq_epsilon_kw_doublesum}
\end{equation}
with complex coefficients $\tilde{\epsilon}_{m,n}$. In the forward-traveling wave described in Eq.~\ref{eq_epsilon_xt}, however, the allowed momentum and frequency transitions are coupled through a 1-to-1 correspondence or, mathematically, 
\begin{equation}
\tilde{\epsilon}_2(k,\omega) = \sum_{n=-\infty}^{\infty} \tilde{\epsilon}_n\delta(\omega-n\Omega) \delta(k+nK).
\label{eq_epsilon_kw_singlesum}
\end{equation}
This leads us to invoke Bloch theorem in the more compact form \cite{PhysRevB.96.155409}
\begin{equation}
E_2(x,t) = e^{j(\omega t - k x)} \sum_{n=-\infty}^{\infty} \tilde{E}_n e^{jn(\Omega t - K x)}.
\label{eq_E_bloch}
\end{equation}
An analogous expansion for $H(x,t)$, together with Eqs.~\ref{eq_epsilon_kw_singlesum},~\ref{eq_E_bloch} into both source-free curl equations \cite{PhysRevA.79.053821}, describes the eigenvalue problem depicting the dispersion diagrams in Figures~\ref{Fig_spacetime_dispersion}(a),(b) (the colorbar identifies the dominating spacetime harmonic order of the associated eigenmodes). Panel (a) represents one of the two sets of parameters in Figure~\ref{fig_temporal_illusion}(c), $Q=-275\frac{{\rm{V}}}{{\rm{m}}}$ and $E_{DC}=3\frac{{\rm{V}}}{{\rm{m}}}$, now with $E_{AC}=0.2\frac{{\rm{V}}}{{\rm{m}}}$. This choice spans a range of $(1.79,14.06)$ for $\epsilon_2$. Keeping the same DC bias with $Q=-151.4851\frac{{\rm{V}}}{{\rm{m}}}$ and $E_{AC}=0.0297\frac{{\rm{V}}}{{\rm{m}}}$ (panel (b)) makes $\epsilon_2$ oscillate between $49$ and $50$. If we first focus on panel (a), we see two dispersion bands coalescing at discrete points along the light cone for $c/\sqrt{\epsilon_1}$ (solid black line), whenever $\omega/\Omega=k/K$ is an integer number. If we take the following limit pairs, $\{\omega,k\} \to \{0^+,0^-\}$ and $\{\omega,k\} \to \{0^+,0^+\}$, and represent the harmonic content of the corresponding joint $\{E,H\}_p$ eigenmode, we obtain in Figure~\ref{Fig_spacetime_dispersion}(b) the solid lines and filled markers, respectively (both denoted by dominant ``order 0''). In the case of the electric field, the solid blue line ($k \to 0^-$) and filled circles ($k \to 0^+$) perfectly overlap and identically describe the denominator in Eqs.~\ref{eq_epsilon_xt} with equal $E_{AC}/E_{DC}$ ratio. This coalescent point is thus the transformed-domain counterpart of making $E_2(x,t)/E_1(x,t)=1$. Remarkably, the magnetic field of these two modes (solid red line ($k \to 0^-$) and filled circles ($k \to 0^+$)) is also identical in the $\pm1$-order harmonic, but has a flip of sign in the magnetostatic component, thus validating our initial \textit{ansatz}: unsurprinsingly, $E_1(x,t)$ is indeed a wave solution in a medium modulated with this very same $E_1(x,t)$ in the denominator of Eq.~\ref{EQ_space_time_permittivity}, and this can be done without the presence of a DC magnetic field. Inspecting the pair of eigenmodes coalescing at $\omega/\Omega=k/K=1$ (dashed lines and void markers, denoted by dominant ``order -1''), we observe identical behavior, except the harmonic indices of $\tilde{E}_n$ and $\tilde{H}_n$ are now shifted by -1 in order to reproduce, through the Bloch expansion, the same fields as before. Of course, commensurate with this tilted bandgap diagram, one cannot find a single-frequency (DC-biased or not) backward wave that is an eigensolution of this medium: this is a clear signature of Lorentz-reciprocity breaking by the modulating forward wave. Any eigenmode other than the replicas of panel (c) along the light cone will necessarily have some other form of ``disordered'' harmonic content, as mandated by the intermodulation products generated in Eq.~\ref{EQ_space_time_permittivity}. 

The general behavior of the medium in panel (b), with $\epsilon_2$ in the range $(49,50)$, is essentially the same. There are two differences, though. (i) We now have a small variation in the dielectric function relative to its average $\langle \epsilon \rangle$, which explains why the diagram is approximately conformed by straight lines, dispersionless-like: all these lines have slopes 
$\pm c / \sqrt{\langle \epsilon \rangle}$ (black dashed lines). (ii) The fact that this average $\langle \epsilon \rangle$ is now closer to $\epsilon_1$ than in panel (a) compresses the diagram along the direction normal to the light cone in the medium: one can think of the limiting case with $\epsilon_2$ in the range $(\epsilon_1-\Delta,\epsilon_1)$, $\Delta$ being infinitesimal; all these compressed lines collapse into a single straight line corresponding to the trivial dispersion of $\epsilon_1$. Regardless, if we zoom in, sufficiently enough, onto any apparent crossing not along the light cone (the discrete set $\omega/\Omega=k/K=n$, with $n$ integer), as in the black circle in the inset of panel (d), we actually find a gap: there is no such coalescence of two modes anymore, and a hint to this fact is given by the different order of the predominant harmonic of the respective eigenmode ($0$ and $1$ in this specific case, \textit{vs.} both $0$'s or both $1$'s in panel (c)). Accordingly, there is no overlap between lines and markers either that can reconstruct $E_1(x,t)$.   

\subsection{Illusion of purely-imaginary-index material, with $\epsilon_1<0$}

\begin{figure*}
\centering
\begin{subfigure}{0.22\textwidth}
\centering
 \includegraphics[width=\textwidth]{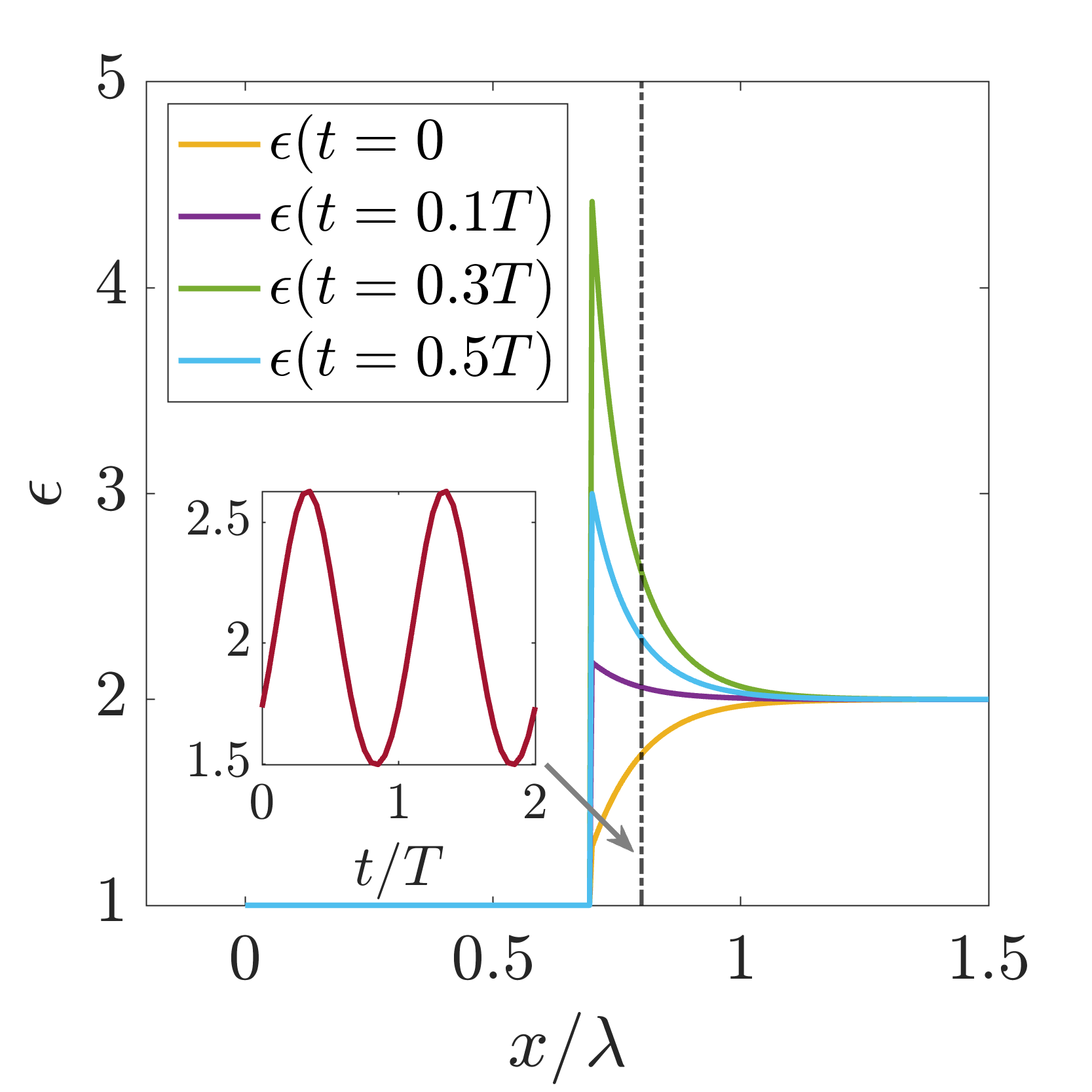}
\caption{}
\end{subfigure}
\begin{subfigure}{0.24\textwidth}
\centering
 \includegraphics[width=\textwidth]{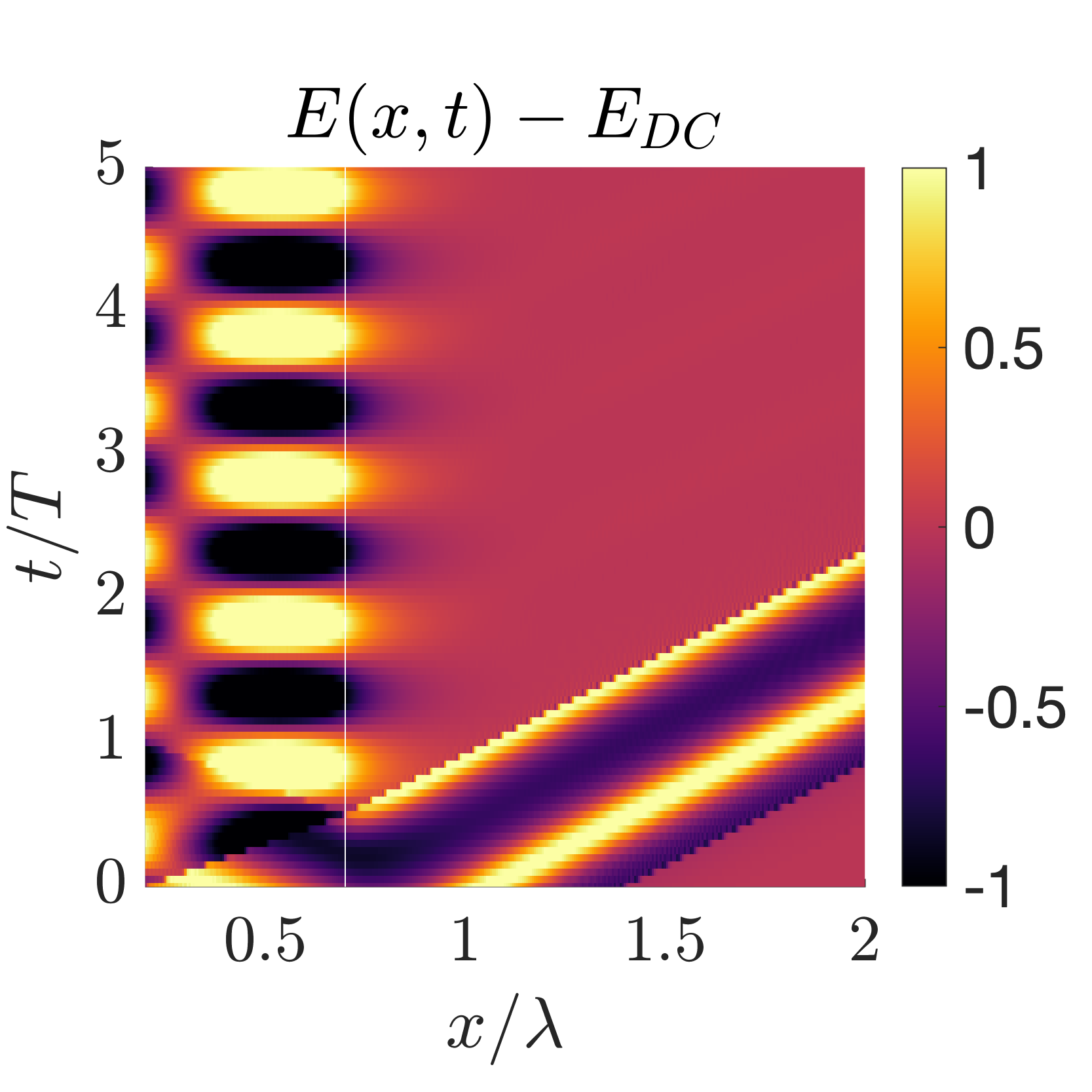}
\caption{}
\end{subfigure}
\begin{subfigure}{0.22\textwidth}
\centering
 \includegraphics[width=\textwidth]{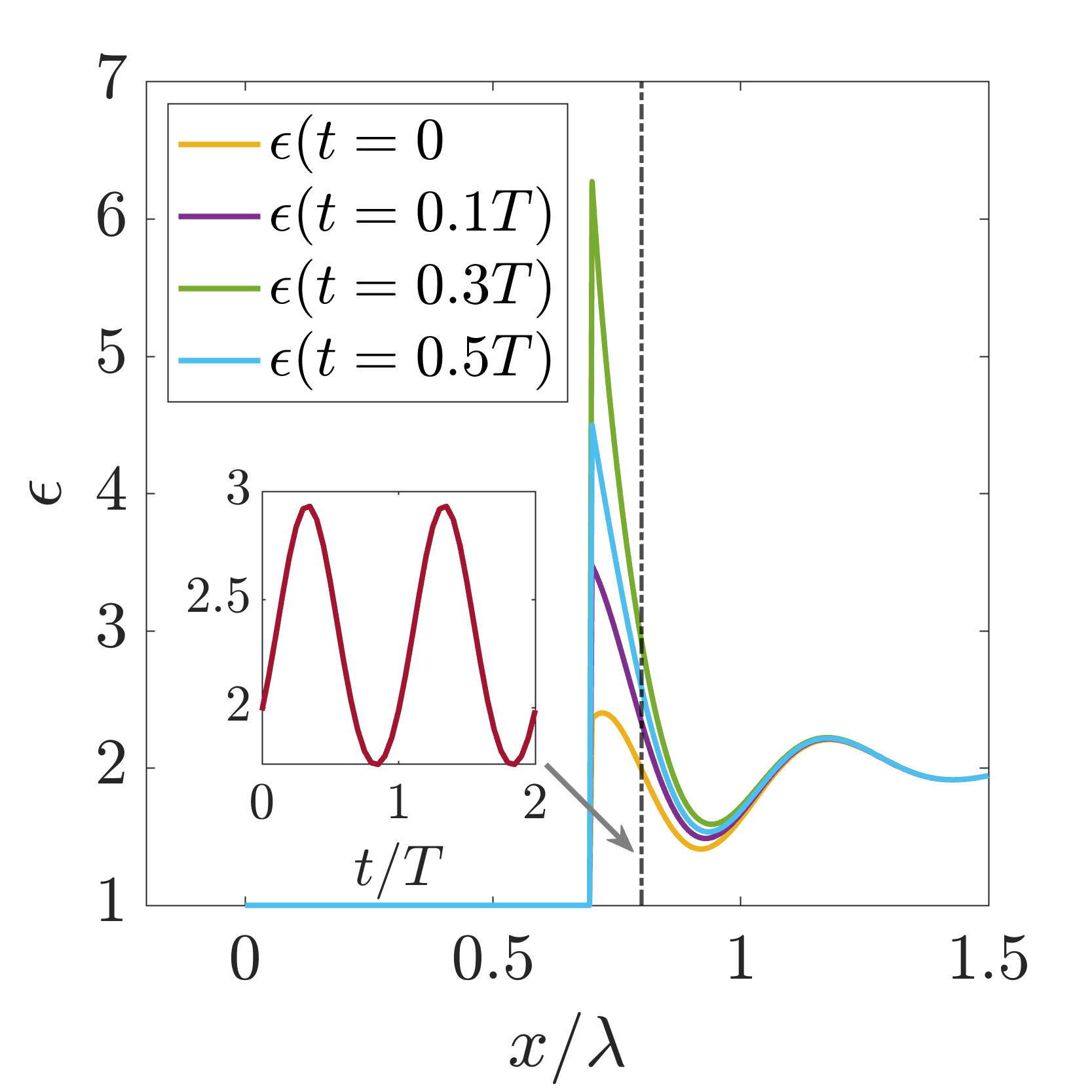}
\caption{}
\end{subfigure}
\begin{subfigure}{0.24\textwidth}
\centering
 \includegraphics[width=\textwidth]{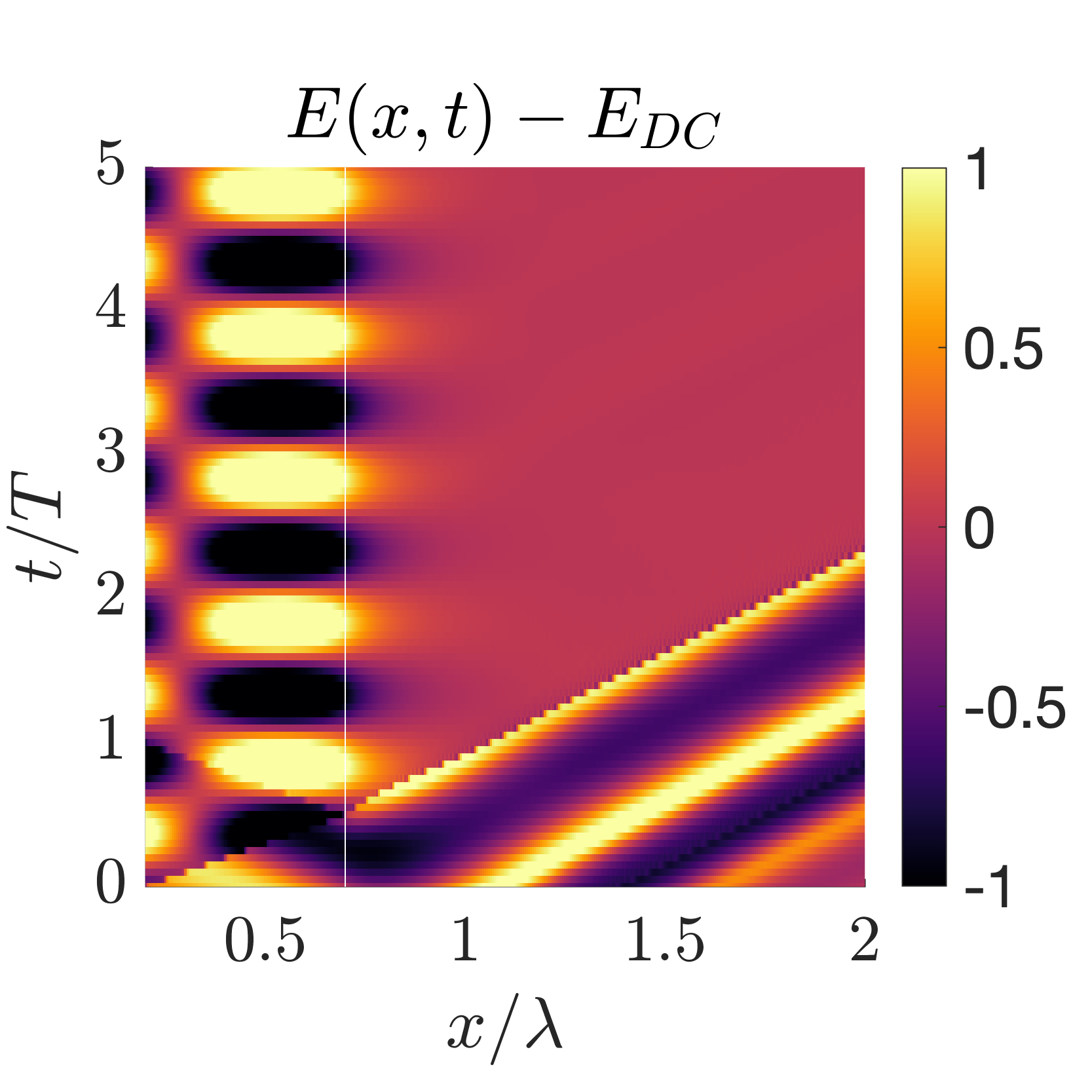}
\caption{}
\end{subfigure}
\\
\begin{subfigure}{0.22\textwidth}
\centering
 \includegraphics[width=\textwidth]{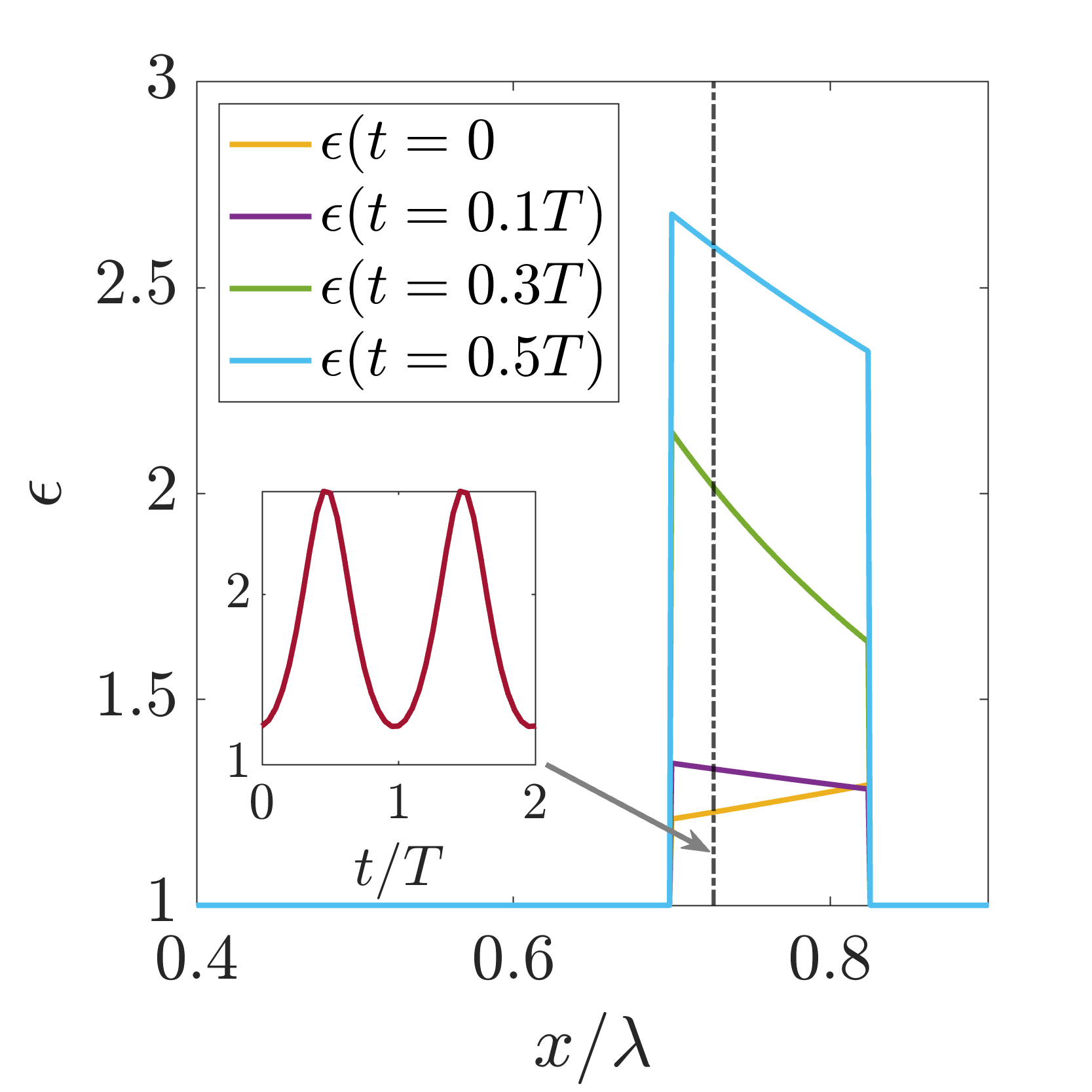}
\caption{}
\end{subfigure}
\begin{subfigure}{0.24\textwidth}
\centering
 \includegraphics[width=\textwidth]{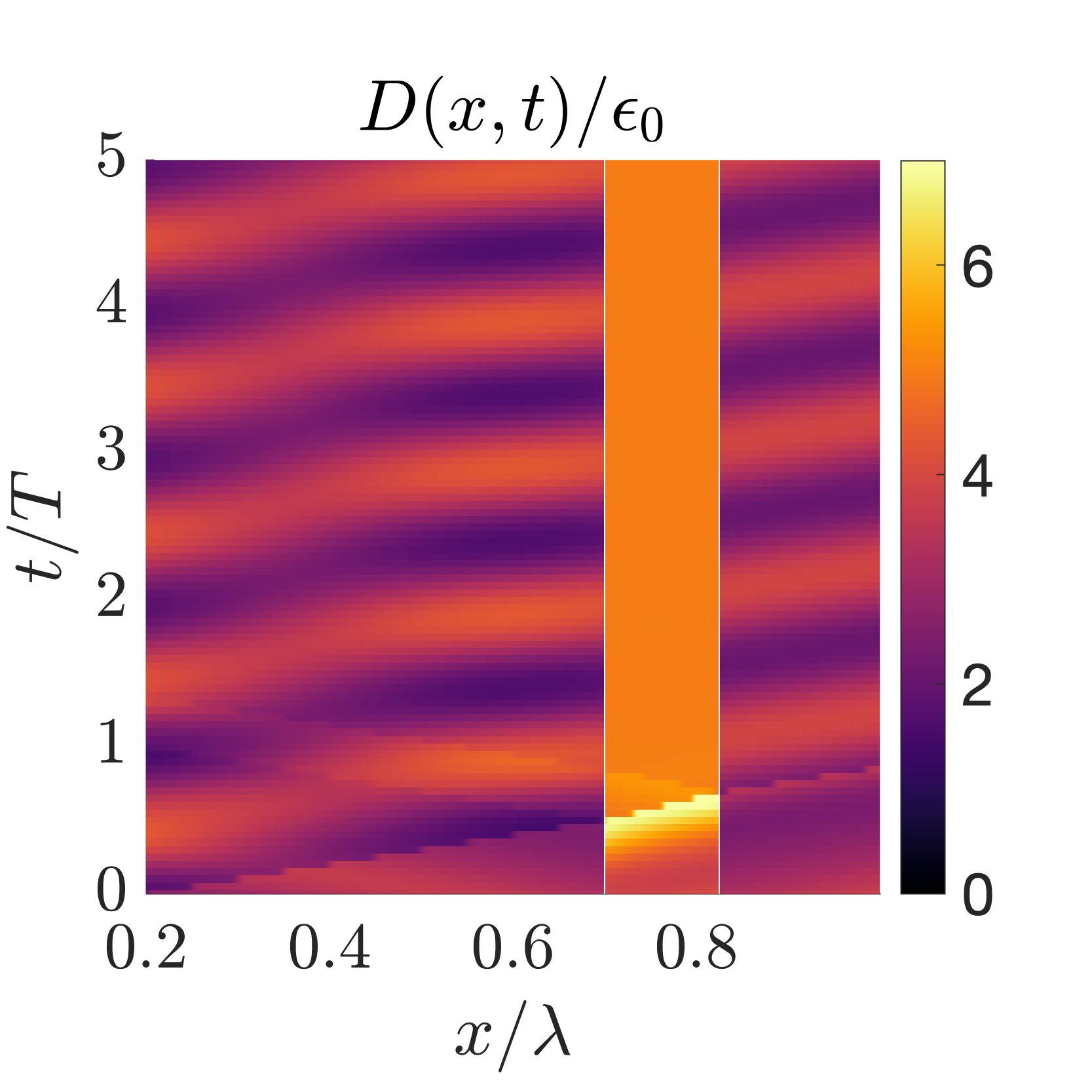}
\caption{}
\end{subfigure}
\centering
\begin{subfigure}{0.24\textwidth}
\centering
 \includegraphics[width=\textwidth]{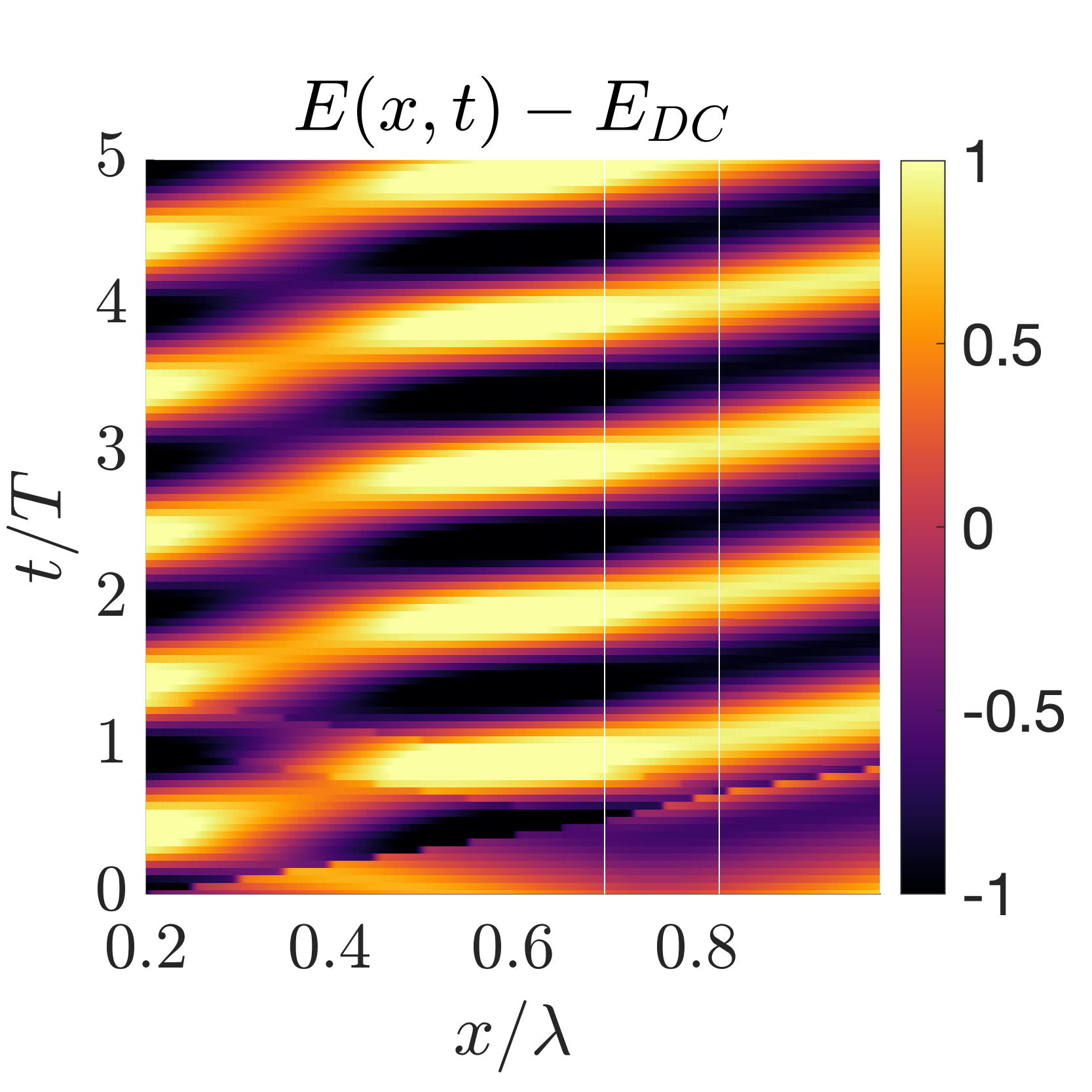}
\caption{}
\end{subfigure}
\begin{subfigure}{0.24\textwidth}
\centering
 \includegraphics[width=\textwidth]{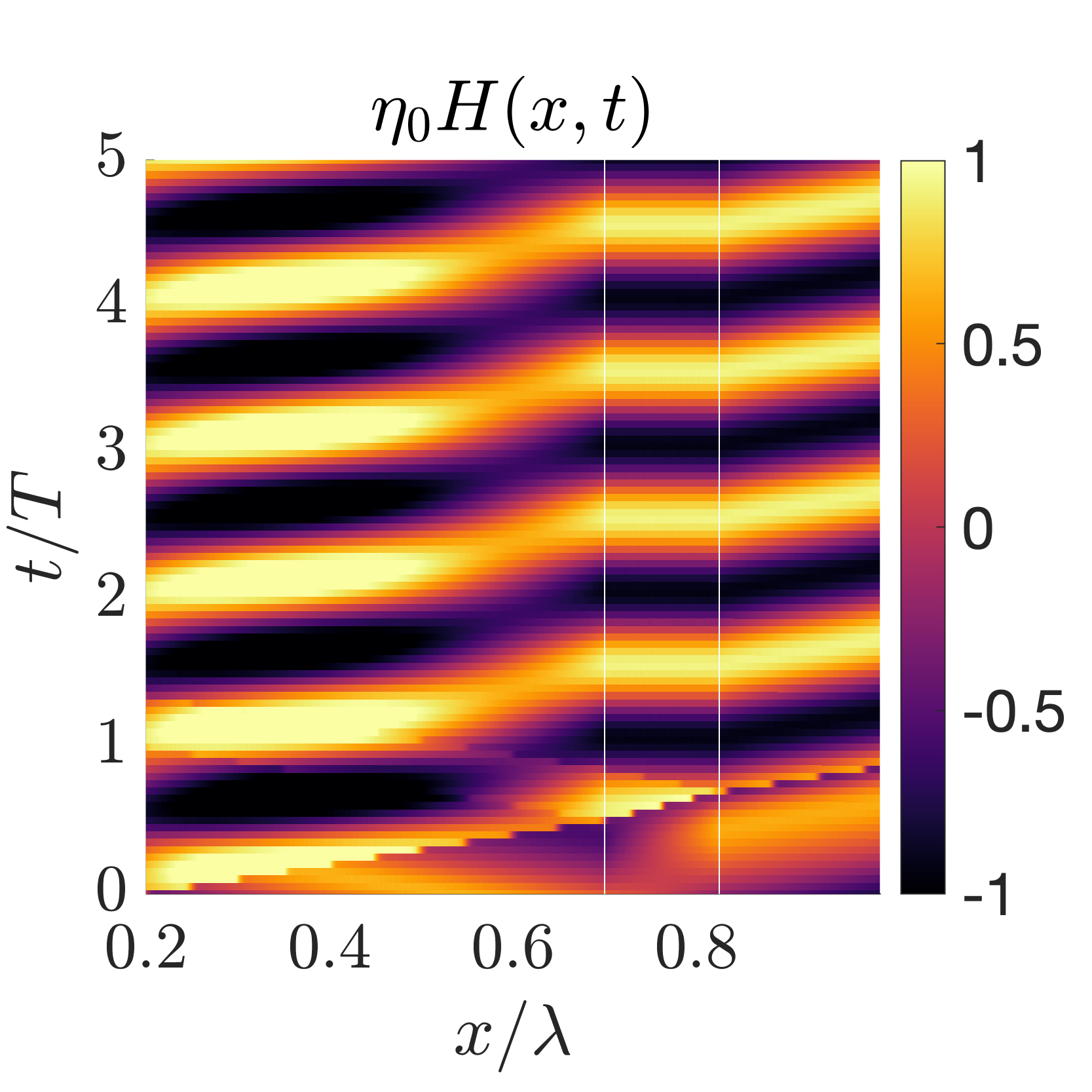}
\caption{}
\end{subfigure}
\caption{\textbf{Temporal illusion for negative-permittivity and epsilon-near-zero media.} 
(a) Space-time varying permittivity of the half-space, $\epsilon_2(x,t)$, that creates an illusion of $\epsilon_1=-3$ (in a Drude medium) for the case where $Q(\_r)=-270\frac{{\rm{V}}}{{\rm{m}}}$ is a constant. 
(b) the electric field in space and time for the cases in (a) shifted by $E_{DC}=3\frac{{\rm{V}}}{{\rm{m}}}$. 
(c)~and~(d) same as in (a) and~(b), respectively, but for the case where $Q(\_r)=Q_0+\frac{Q_0}{4}\exp{(-\frac{\alpha}{3}\frac{\omega_0}{c}x)}\cos{(2\frac{\omega_0}{c}x)}$, where $Q_0=-270 \frac{\rm V}{\rm m}$. 
(e) space-time varying $\epsilon_2(x,t)$ that creates an illusion of slab (thickness $d=\lambda/8$) made of ENZ $\epsilon_1=10^{-5}$ medium, where $Q(\_r)=5\frac{{\rm{V}}}{{\rm{m}}}$ is a constant.
(f),~(g) and~(h) Space-time distribution of normalized electric flux density, shifted by $E_{DC}=3\frac{{\rm{V}}}{{\rm{m}}}$ electric field and normalized magnetic field, respectively. Alternating part of the electric field $E_{AC}=1\frac{{\rm{V}}}{{\rm{m}}}$ for all cases.} 

\label{fig_temporal_illusion2}
\end{figure*} 

\noindent In this subsection, we demonstrate an illusion of a medium characterized by negative permittivity. If we assume that this negative permittivity is dispersionless, i.e., it is of a non-Foster kind, the feedback modulation scheme cannot be used here because a medium with nondispersive negative permittivity essentially violates Foster's reactance theorem~\cite{foster1924reactance}, which implies that it should be an active medium as it amplifies the fields exponentially in time. Therefore, to avoid this scenario, we consider a passive medium with the Drude dispersion.  In particular, we consider a half-space (not a finite-thickness slab) with permittivity given by the Drude dispersion, $\epsilon_1(\omega)=1-\omega_{\rm{p}}^2/\omega^2$. We select the plasma frequency $\omega_{\rm p}$ to be greater than the excitation frequency $\omega_0$, so that $\epsilon_1=-3$. Due to the negative permittivity, the wave number in the half space becomes purely imaginary, resulting in an exponential decay of the wave in the Drude material. Furthermore, since the wave impedance is also purely imaginary, this decaying field does not transport active time-average power, since the time-average Poynting vector is zero. Consequently, all incident power is totally reflected back at the interface.  With this information on the scattering, one can derive the required space-time-varying permittivity for $\epsilon_2(x,t)$, which always stays positive. (see Supplementary Materials for derivations). 

Figure~\ref{fig_temporal_illusion2}(a) shows the space-time-varying permittivity of the half space for the case when $Q(\_r)=-270\frac{{\rm{V}}}{{\rm{m}}}$ is a constant. 
Figures~\ref{fig_temporal_illusion2}(b) shows the corresponding electric field.
One can notice that the electric field decays inside the medium after $1T$, except for the initial transient disturbance that propagates into the half-space, which can be attributed to radiation from the slab occurring due to temporally varying permittivity in presence of the DC electric field. 
Once the incident wave reaches the half-space boundary, it cancels out this radiation, making the total time-varying field exponentially dropping inside the medium.

As noted above, the parameter $Q$ can be arbitrarily chosen to satisfy the conditions of $\epsilon_2(\_r,t)$, thereby providing control over the transition period. However, there is another interesting feature for $Q(\_r)$, as it does not necessarily need to be uniform in space; it can be defined as an arbitrary spatial function, but temporally time-invariant, thus enabling the engineering of the spatial modulation of $\epsilon_2(\_r,t)$. Figure~\ref{fig_temporal_illusion2}(c) illustrates a modulation profile that creates the illusion of $\epsilon_1=-3$, obtained by selecting $Q(\_r)=Q_0+\frac{Q_0}{4}\exp{(-\frac{\alpha}{3}\frac{\omega_0}{c}x)}\cos{(2\frac{\omega_0}{c}x)}$, where $Q_0=-270 \frac{\rm V}{\rm m}$. The corresponding electric-field dynamics in space and time are shown in Fig.~\ref{fig_temporal_illusion2}(d). It can be observed that, after the transient period ($1T$), the electric field does drop exponentially into the half-space in both cases; however, the transient periods slightly differ. 

\subsection{Illusion of near-zero-index material, with $\epsilon_1 \approx 0$}

\noindent The concept of temporal illusion can be applied for realization of epsilon-near-zero (ENZ) media at an arbitrary frequency. A nondispersive medium with a permittivity smaller than unity is essentially non-Foster as the corresponding susceptibility is negative. Therefore, analogously to negative permittivity, for an illusion of epsilon-near-zero medium, the instantaneous feedback modulation scheme cannot be applied. However, the steady-state modulation can be employed as the steady-state solution can be analytically calculated using the same procedure mentioned above for a slab with arbitrarily large $\epsilon_1$ (see Supplementary information).

In this subsection, we consider a slab made of  material with relative permittivity positive and near zero, e.g., $\epsilon_1=10^{-5}$ at a given frequency. Figure~\ref{fig_temporal_illusion2}(e) presents the modulation function $\epsilon_2(x,t)$, which varies with $x$ at any moment in time. The corresponding simulation results are shown in Figs.~\ref{fig_temporal_illusion2}(f),~(g), and~(h). Specifically, Fig.~\ref{fig_temporal_illusion2}(f) shows normalized electric flux density [$\_D(x,t)/\epsilon_0$], while Figs.~\ref{fig_temporal_illusion2}(g) and~(h) depict the electric field shifted by $E_{DC}$ and the magnetic field scaled by $\eta_0$, respectively. 

Inside an actual ENZ medium, electric flux density $\_D$ remains close to zero, because the permittivity is near zero. In our case, this quantity ($\_D$) is not zero and, in fact, is constant across the slab. This is explained by the presence of the DC electric field.  Notably, the absence of time-varying part of $\_D$ is a consequence of a completely different mechanism from that in the actual ENZ medium: In our case, a combination of linearly changing in space electric field $\_E(x,t)$ and $\epsilon_2(x,t)$ cancels the time-varying component of $\_D$. 
Particularly, at any given moment of time, one of them is linearly increasing with $x$ and the other is linearly decreasing with $x$, so when they are multiplied at each point of space, it gives a constant value. As a consequence of a linearly varying electric field in space, following the Faraday law the time-derivative of the magnetic field is uniform across the slab, which is consistent with ENZ behavior, causing the magnetic flux density to be a function of time only.

As we have observed thus far, at a given frequency, temporal illusion could replicate the macroscopic responses of time-invariant one-dimensional structures, provided that the time-varying system under study is modulated following the expression we discussed above. This requirement highlights a salient feature closely tied to the notion of detuning, which we elaborate upon in the following.

\subsection{Introducing detuning between modulation and excitation}

\noindent The concept of temporal illusion requires perfect synchronization between modulation and excitation.
In case of instantaneous feedback modulation scheme, the ideal synchronization is assumed and detuning would mean introducing a temporal delay to modulation. 
Although it is possible, this is not discuss here, as the steady state modulation scenario is central in our work here.

\begin{figure*}
\begin{subfigure}{0.35\textwidth}
\centering
 \includegraphics[width=\textwidth]{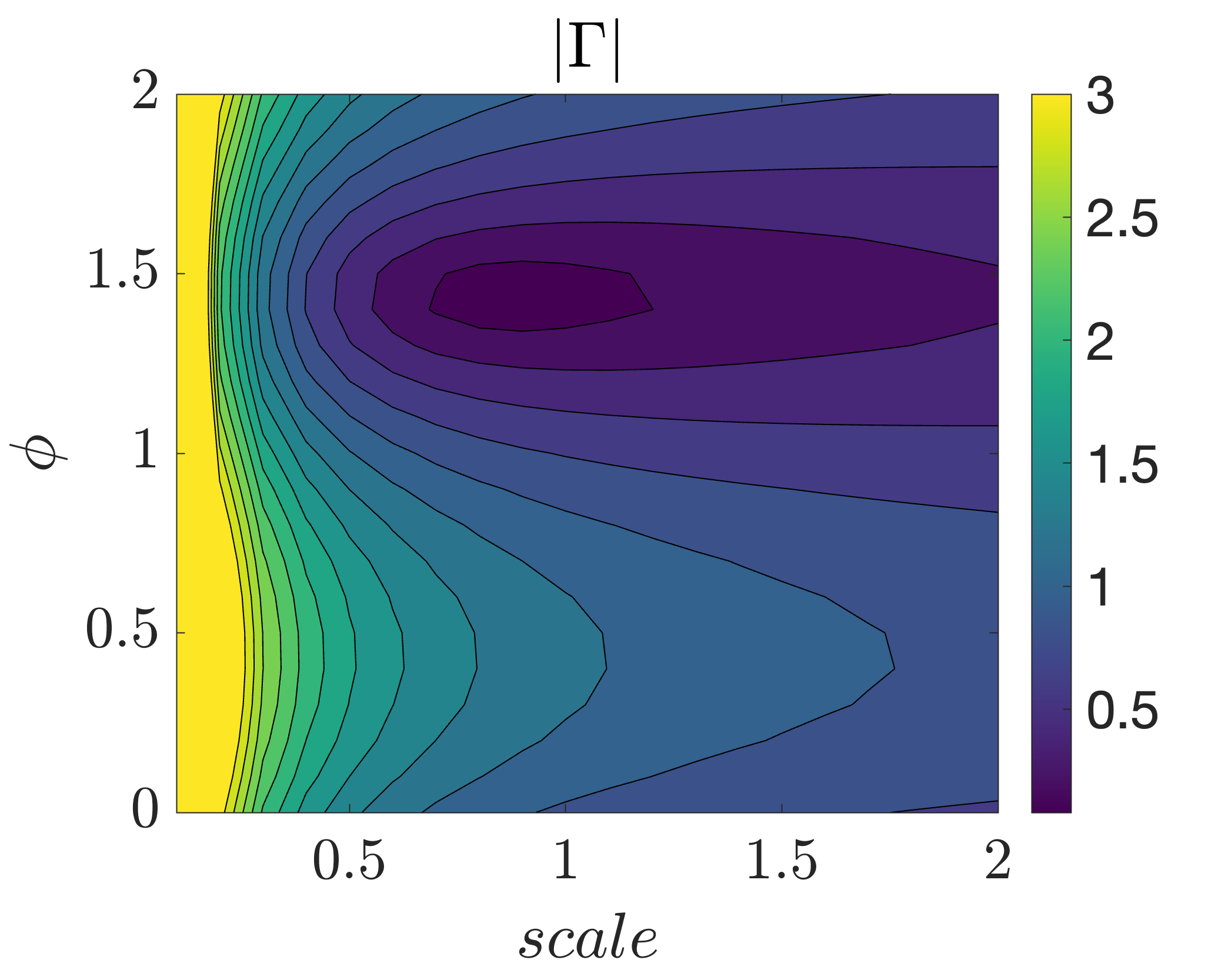}
\caption{}
\end{subfigure}
~
\begin{subfigure}{0.35\textwidth}
\centering
 \includegraphics[width=\textwidth]{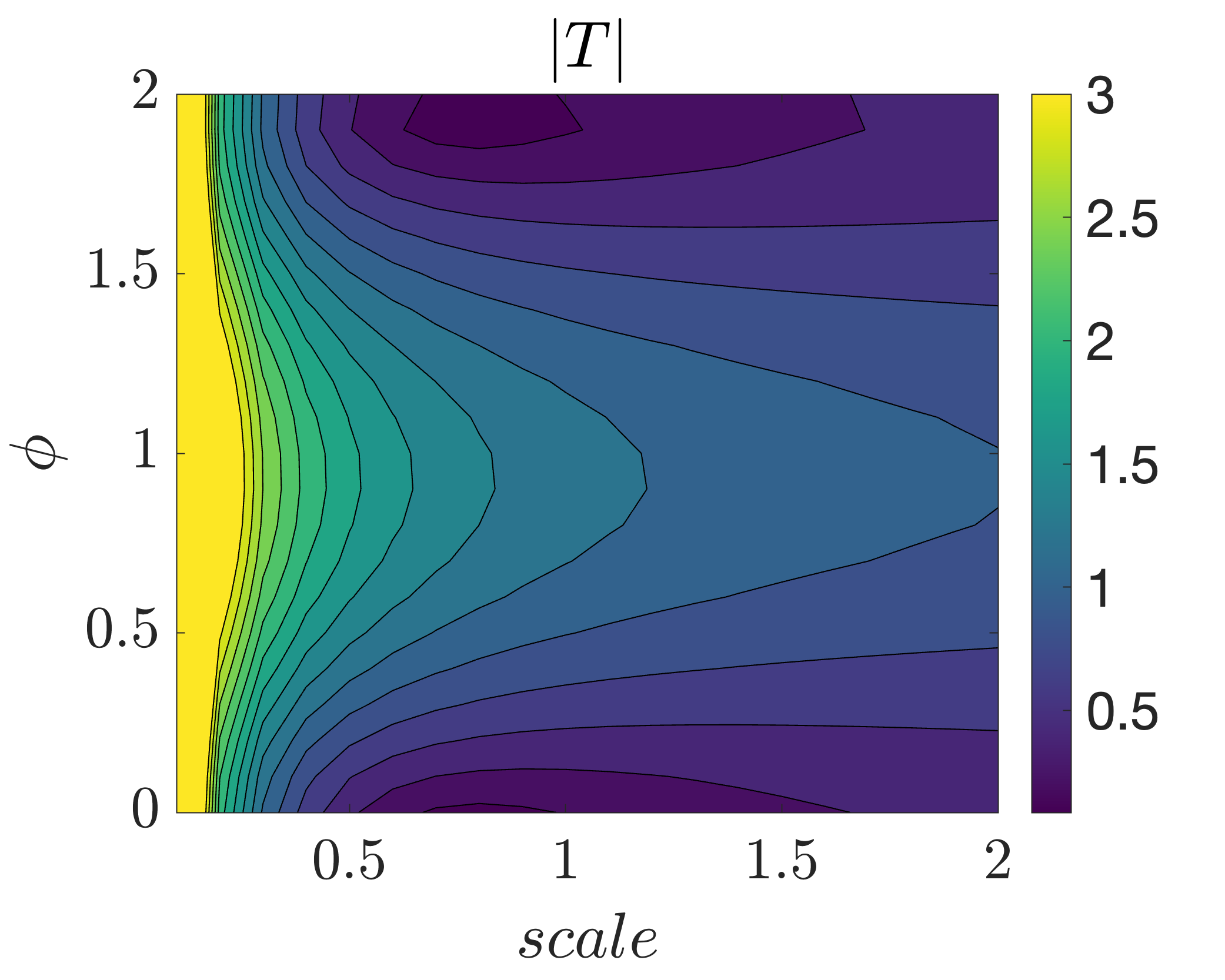}
\caption{}
\end{subfigure}
\\
\begin{subfigure}{0.35\textwidth}
\centering
 \includegraphics[width=\textwidth]{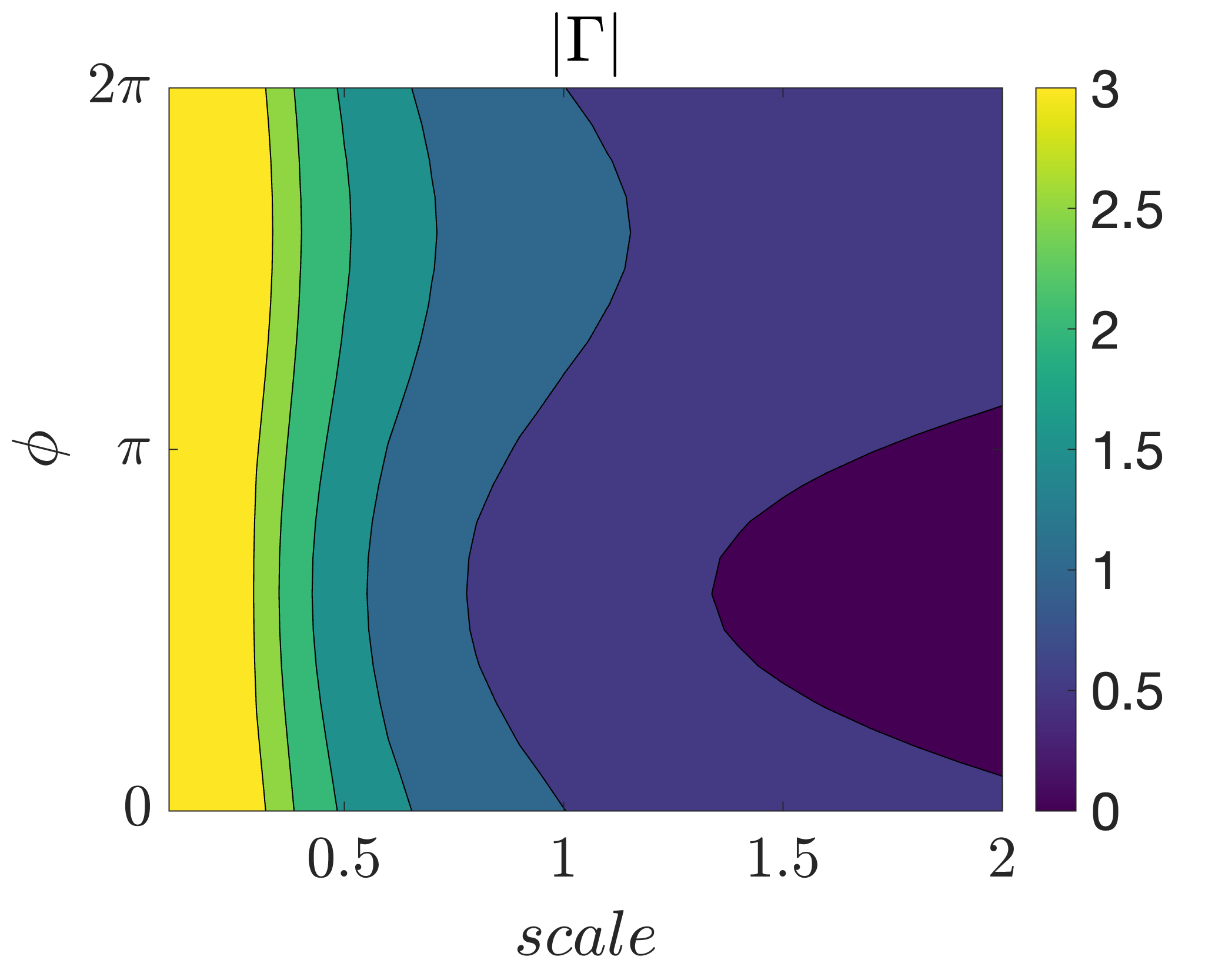}
\caption{}
\end{subfigure}
~
\begin{subfigure}{0.35\textwidth}
\centering
 \includegraphics[width=\textwidth]{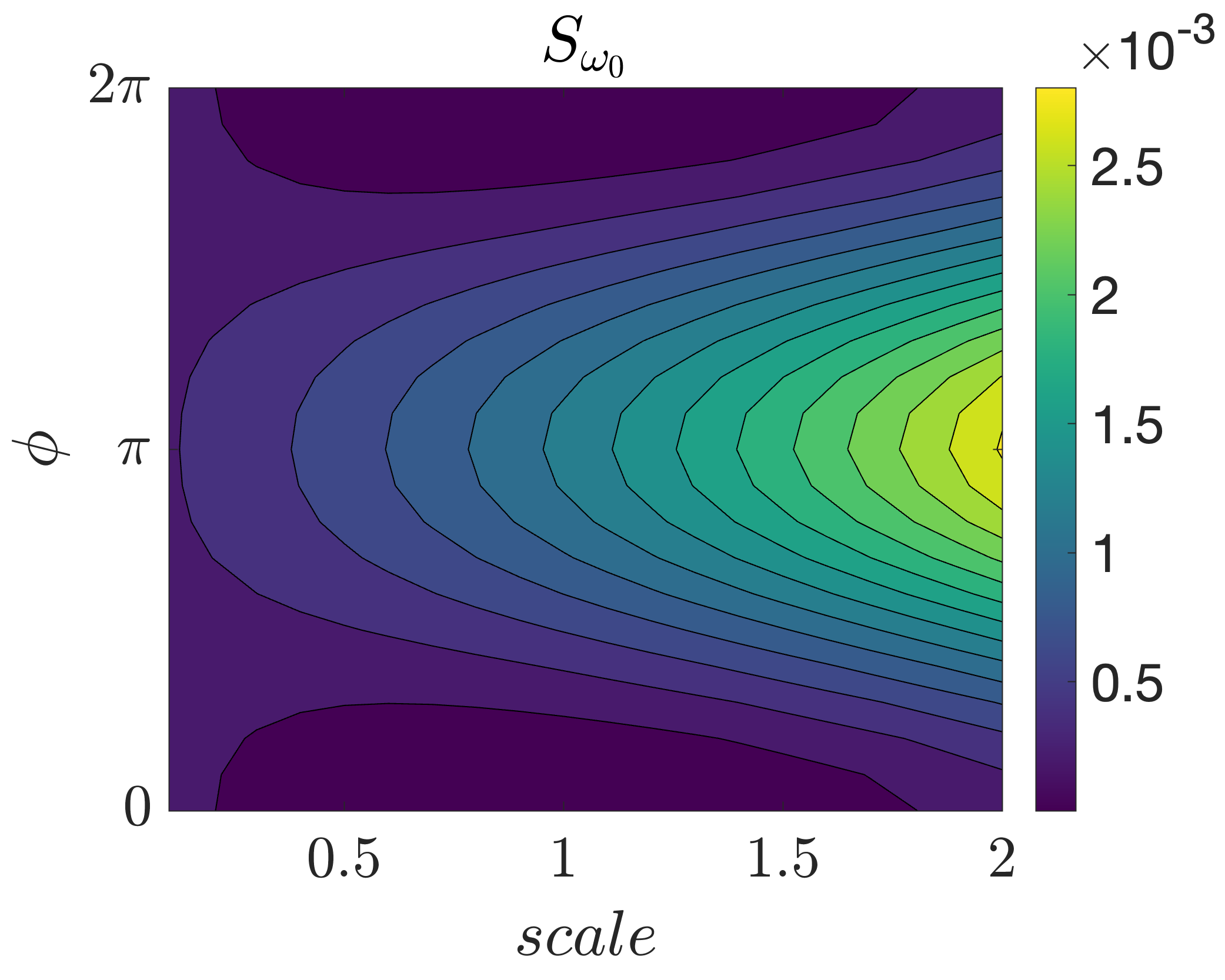}
\caption{}
\end{subfigure}
\caption{\textbf{Detuning between excitation and steady-state modulation.} Magnitudes of the reflection and transmission coefficients in~(a) and~(b), respectively, for space-time varying slab that creates an illusion of $\epsilon_1=100$ when scale and $\phi$ are varied. (c) and~(d) Space-time varying half-space that in the synchronized case creates an illusion of $\epsilon_1=-3$. (c) shows the magnitude of the reflection coefficient and (d) shows the time-average Poynting vector at distance $\lambda$ inside the space-time varying medium when scale and phase $\phi$ of excitation are detuned. }
\label{fig_detuning}
\end{figure*}

For steady state modulation scheme the modulation function is precalculated and the detuning will be introduced by a mismatch in phase and amplitude. 
Naturally, detuning leads to a different response from the same structure. 
For simplicity, in this section the modulation function is fixed, whereas the relative amplitude (which we  denote as "scale")  and phase (denoted as $\phi$)  of the incident signal are varied. Two cases, shown in the previous section, are considered further here: the space-time varying slab that, in the synchronized case, creates an illusion of $\epsilon_1=100$ and the space-time varying half-space that mimics negative $\epsilon_1=-3$. 
Further, $Q(\_r)$ is assumed uniform in space, and the modulation functions for these cases can be found in Figs.~\ref{fig_temporal_illusion}(c) and~\ref{fig_temporal_illusion2}(a). 
Figure~\ref{fig_detuning}(a) and~(b) show reflection and transmission coefficients for the excitation frequency $\omega_0$. 
For the calculation of $|\Gamma|$ and $|T|$ the electric field was taken after a long enough period of time, i.e. the steady state.
One can notice that these coefficients can be engineered in a wide range, providing amplification or attenuation for either of them. 

Figure~\ref{fig_detuning}(c) shows reflection coefficient for a half-space problem and Fig.~\ref{fig_detuning}(d) shows Poynting vector deep inside ($1\lambda$) the space-time varying medium. 
One can see that for the synchronized case ($scale=1$ and $\phi=0$) reflection coefficient $|\Gamma|=1$ and Poynting vector at the frequency of excitation ($\omega_0$) $S_{\omega_0}$ is equal to 0, indicating that all the incident energy is reflected and there is no power flow into the medium.
However, broken synchronization enables control over the reflection coefficient and the power flow.  

Interestingly, the concept of temporal illusion involves time modulation at the frequency of excitation ($\omega_0$), which inevitably excites higher-order harmonics, i.e. $n\omega_2$, where $n$ is an arbitrary integer number. 
However, in the synchronized case, the spectral content of the fields after the transition time consists almost exclusively of the main harmonic ($\omega_0$).
However, when the synchronization is broken, we will obtain higher-order Floquet harmonics. 


\section{Summary}

\noindent In this work, we have introduced the concept of photonic temporal illusion, which enables mimicking response of time-invariant structures with arbitrary permittivity.
Particularly, we showed how to create an illusion of a slab with arbitrary positive permittivity, a slab with near-zero refractive index, and a half-space with Drude dispersion with arbitrary negative permittivity. 
We discussed the benefits of temporal illusion and highlighted that one can control the time it takes for the system to converge to the steady state, which has potential use in high-quality factor cavities.  We have also discussed the effects of detuning/desynchronization between the modulating the permittivity and the incident signal. We have also conducted full wave simulations using customized FDTD code that corroborates our findings. 
Future development of temporal illusion concept can follow several paths, including transfer of this idea to acoustics, thermal radiation, and any other wave based physical system.
Additionally, we envisage extension of temporal illusion to 2D and 3D problems, which enables mimicking various physical phenomena including plasmonic resonances, surface plasmon polariton and more. 






\begin{acknowledgments}
G.P. acknowledges the support from Ulla Tuominen Foundation. M.S.M.~gratefully acknowledges the support from the Research Council of Finland (Grant No.~336119). D.M.S acknowledges support from MICIU/AEI/10.13039 /501100011033 and ESF+ (Ramón y Cajal fellowship: grant RYC2023-045265-I), and Xunta de Galicia (Consolidation of Competitive Research Units type C: grant ED431F 2025/22).  N.E. acknowledges the support from the Simons
Foundation/Collaboration on Symmetry-Driven Extreme
Wave Phenomena (grant SFI-MPSEWP-00008530-04).
\end{acknowledgments} 


\bigskip

\noindent{\textbf{Disclosures.}}~The authors declare no conflicts of interest.

\noindent{\textbf{Data availability.}}~Data underlying the results presented in this paper are not publicly available at this time but may be obtained from the authors upon reasonable request.


\bibliography{references1}

@article{engheta2020metamaterials,
  title={Metamaterials with high degrees of freedom: space, time, and more},
  author={Engheta, Nader},
  journal={Nanophotonics},
  volume={10},
  number={1},
  pages={639--642},
  year={2020},
  publisher={De Gruyter}
}

@article{foster1924reactance,
  title={A Reactance Theorem},
  author={Foster, Ronald M},
  journal={Bell System technical journal},
  volume={3},
  number={2},
  pages={259--267},
  year={1924},
  publisher={Wiley Online Library}
}

@article{engheta2023four,
  title={Four-dimensional optics using time-varying metamaterials},
  author={Engheta, Nader},
  journal={Science},
  volume={379},
  number={6638},
  pages={1190--1191},
  year={2023},
  publisher={American Association for the Advancement of Science}
}

@article{galiffi_photonics_2022,
  title = {Photonics of Time-Varying Media},
  author = {Galiffi, Emanuele and Tirole, Romain and Yin, Shixiong and Li, Huanan and Vezzoli, Stefano and Huidobro, Paloma A. and Silveirinha, M{\'a}rio G. and Sapienza, Riccardo and Al{\`u}, Andrea and Pendry, J. B.},
  year = {2022},
  month = feb,
  journal = {Advanced Photonics},
  volume = {4},
  number = {1},
  pages = {014002},
  publisher = {{SPIE}},
  issn = {2577-5421, 2577-5421},
  doi = {10.1117/1.AP.4.1.014002}
}

@article{zhou2020broadband,
  title={Broadband frequency translation through time refraction in an epsilon-near-zero material},
  author={Zhou, Yiyu and Alam, M Zahirul and Karimi, Mohammad and Upham, Jeremy and Reshef, Orad and Liu, Cong and Willner, Alan E and Boyd, Robert W},
  journal={Nature Communications},
  volume={11},
  number={1},
  pages={2180},
  year={2020},
  publisher={Nature Publishing Group}
}

@article{pacheco2020temporal,
  title={Temporal aiming},
  author={Pacheco-Pe{\~n}a, Victor and Engheta, Nader},
  journal={Light: Science \& Applications},
  volume={9},
  number={1},
  pages={129},
  year={2020},
  publisher={Nature Publishing Group UK London}
}

@article{yin2022efficient,
  title={Efficient phase conjugation in a space-time leaky waveguide},
  author={Yin, Shixiong and Al{\`u}, Andrea},
  journal={ACS Photonics},
  volume={9},
  number={3},
  pages={979--984},
  year={2022},
  publisher={ACS Publications}
}

@article{quinones_tunable_2021,
  title = {Tunable Surface Plasmon Resonance in Laser-Induced Plasma Spheroids},
  author = {Qui{\~n}ones, Roberto A. Col{\'o}n and Underwood, Thomas Carlton and Cappelli, Mark A.},
  year = {2021},
  month = apr,
  journal = {Plasma Sources Science and Technology},
  volume = {30},
  number = {4},
  pages = {045010},
  publisher = {{IOP Publishing}},
  issn = {0963-0252},
  doi = {10.1088/1361-6595/abc5a2}
}

@article{biancalana_dynamics_2007,
  author   = {Biancalana, Fabio and Amann, Andreas and Uskov, Alexander V. and O'Reilly, Eoin P.},
  title    = {Dynamics of light propagation in spatiotemporal dielectric structures},
  doi      = {10.1103/PhysRevE.75.046607},
  number   = {4},
  pages    = {046607},
  volume   = {75},
  journal  = {Physical Review E},
  year     = {2007}
}

@Article{reyes-ayona_observation_2015,
  author  = {Reyes-Ayona, J. R. and Halevi, P.},
  title   = {Observation of genuine wave vector (k or $\beta$) gap in a dynamic transmission line and temporal photonic crystals},
  doi     = {10.1063/1.4928659},
  issn    = {0003-6951},
  number  = {7},
  pages   = {074101},
  url     = {https://aip.scitation.org/doi/abs/10.1063/1.4928659},
  urldate = {2021-12-03},
  volume  = {107},
  journal = {Applied Physics Letters},
  month   = aug,
  year    = {2015},
}

@article{zurita-sanchez_reflection_2009,
  author   = {Zurita-S\'{a}nchez, Jorge R. and Halevi, P. and Cervantes-Gonzalez, Juan C.},
  title    = {Reflection and transmission of a wave incident on a slab with a time-periodic dielectric function $\varepsilon(t)$},
  doi      = {10.1103/PhysRevA.79.053821},
  number   = {5},
  pages    = {053821},
  url      = {https://link.aps.org/doi/10.1103/PhysRevA.79.053821},
  urldate  = {2021-12-03},
  volume   = {79},
  journal  = {Physical Review A},
  month    = may,
  year     = {2009}
}

@article{lustig_topological_2018,
	title = {Topological aspects of photonic time crystals},
	volume = {5},
	copyright = {\&\#169; 2018 Optical Society of America},
	issn = {2334-2536},
	url = {https://www.osapublishing.org/optica/abstract.cfm?uri=optica-5-11-1390},
	number = {11},
	urldate = {2021-12-03},
	journal = {Optica},
	author = {Lustig, Eran and Sharabi, Yonatan and Segev, Mordechai},
	month = nov,
	year = {2018},
	pages = {1390--1395},
}

@Article{park_spatiotemporal_2021,
  author    = {Park, Jagang and Min, Bumki},
  title     = {Spatiotemporal plane wave expansion method for arbitrary space--time periodic photonic media},
  doi       = {10.1364/OL.411622},
  issn      = {1539-4794},
  number    = {3},
  pages     = {484--487},
  journal   = {Optics Letters},
  month     = feb,
  year      = {2021},
}

@article{sharabi2021disordered,
  title={Disordered Photonic Time Crystals},
  author={Sharabi, Yonatan and Lustig, Eran and Segev, Mordechai},
  journal={Physical Review Letters},
  volume={126},
  number={16},
  pages={163902},
  year={2021},
  publisher={APS}
}

@article{GregAtom,
	title={Time-Modulated Meta-Atoms},
	author={Ptitcyn, GA and Mirmoosa, MS and Tretyakov, SA},
	journal={Physical Review Research},
	volume={1},
	number={2},
	pages={023014},
	year={2019},
	publisher={APS}
}

@article{fang_realizing_2012,
	title = {Realizing effective magnetic field for photons by controlling the phase of dynamic modulation},
	volume = {6},
	copyright = {2012 Nature Publishing Group},
	issn = {1749-4893},
	number = {11},
	urldate = {2021-12-05},
	journal = {Nature Photonics},
	author = {Fang, Kejie and Yu, Zongfu and Fan, Shanhui},
	month = nov,
	year = {2012},
}

@article{vezzoli_optical_2018,
	title = {Optical time reversal from time-dependent epsilon-near-zero media},
	volume = {120},
	number = {4},
	urldate = {2021-12-03},
	journal = {Physical Review Letters},
	author = {Vezzoli, Stefano and Bruno, Vincenzo and DeVault, Clayton and Roger, Thomas and Shalaev, Vladimir M. and Boltasseva, Alexandra and Ferrera, Marcello and Clerici, Matteo and Dubietis, Audrius and Faccio, Daniele},
	month = jan,
	year = {2018},
	pages = {043902},
}

@article{salary2018time,
  title={Time-varying metamaterials based on graphene-wrapped microwires: Modeling and potential applications},
  author={Salary, Mohammad Mahdi and Jafar-Zanjani, Samad and Mosallaei, Hossein},
  journal={Physical Review B},
  volume={97},
  number={11},
  pages={115421},
  year={2018},
  publisher={APS}
}

@article{wang2018photonic,
  title={Photonic Floquet media with a complex time-periodic permittivity},
  author={Wang, Neng and Zhang, Zhao-Qing and Chan, Che Ting},
  journal={Physical Review B},
  volume={98},
  number={8},
  pages={085142},
  year={2018},
  publisher={APS}
}

@article{koutserimpas2018nonreciprocal,
  title={Nonreciprocal gain in non-Hermitian time-Floquet systems},
  author={Koutserimpas, Theodoros T and Fleury, Romain},
  journal={Physical Review Letters},
  volume={120},
  number={8},
  pages={087401},
  year={2018},
  publisher={APS}
}

@article{ramaccia2017doppler,
  title={Doppler cloak restores invisibility to objects in relativistic motion},
  author={Ramaccia, Davide and Sounas, Dimitrios L and Al{\`u}, Andrea and Toscano, Alessandro and Bilotti, Filiberto},
  journal={Physical Review B},
  volume={95},
  number={7},
  pages={075113},
  year={2017},
  publisher={APS}
}

@article{ramaccia2019phase,
  title={Phase-induced frequency conversion and doppler effect with time-modulated metasurfaces},
  author={Ramaccia, Davide and Sounas, Dimitrios L and Al{\`u}, Andrea and Toscano, Alessandro and Bilotti, Filiberto},
  journal={IEEE Transactions on Antennas and Propagation},
  volume={68},
  number={3},
  pages={1607--1617},
  year={2019},
  publisher={IEEE}
}

@article{huidobro2019fresnel,
  title={Fresnel drag in space--time-modulated metamaterials},
  author={Huidobro, Paloma A and Galiffi, Emanuele and Guenneau, S{\'e}bastien and Craster, Richard V and Pendry, JB},
  journal={Proceedings of the National Academy of Sciences},
  volume={116},
  number={50},
  pages={24943--24948},
  year={2019},
  publisher={National Acad Sciences}
}

@article{liu2019time,
  title={Time-varying metasurfaces for broadband spectral camouflage},
  author={Liu, Mingkai and Kozyrev, Alexander B and Shadrivov, Ilya V},
  journal={Physical Review Applied},
  volume={12},
  number={5},
  pages={054052},
  year={2019},
  publisher={APS}
}

@article{wang2020spread,
  title={Spread-spectrum selective camouflaging based on time-modulated metasurface},
  author={Wang, Xiaoyi and Caloz, Christophe},
  journal={IEEE Transactions on Antennas and Propagation},
  volume={69},
  number={1},
  pages={286--295},
  year={2020},
  publisher={IEEE}
}

@article{yu2009complete,
  title={Complete optical isolation created by indirect interband photonic transitions},
  author={Yu, Zongfu and Fan, Shanhui},
  journal={Nature Photonics},
  volume={3},
  number={2},
  pages={91--94},
  year={2009},
  publisher={Nature Publishing Group}
}

@article{sounas2014angular,
  title={Angular-momentum-biased nanorings to realize magnetic-free integrated optical isolation},
  author={Sounas, Dimitrios L and Al{\`u}, Andrea},
  journal={ACS Photonics},
  volume={1},
  number={3},
  pages={198--204},
  year={2014},
  publisher={ACS Publications}
}

@article{shi2017optical,
  title={Optical circulation and isolation based on indirect photonic transitions of guided resonance modes},
  author={Shi, Yu and Han, Seunghoon and Fan, Shanhui},
  journal={ACS Photonics},
  volume={4},
  number={7},
  pages={1639--1645},
  year={2017},
  publisher={ACS Publications}
}

@article{dinc2017synchronized,
  title={Synchronized conductivity modulation to realize broadband lossless magnetic-free non-reciprocity},
  author={Dinc, Tolga and Tymchenko, Mykhailo and Nagulu, Aravind and Sounas, Dimitrios and Al{\`u}, Andrea and Krishnaswamy, Harish},
  journal={Nature Communications},
  volume={8},
  number={1},
  pages={795},
  year={2017},
  publisher={Nature Publishing Group}
}

@article{Our2,
  title={Nonreciprocity in bianisotropic systems with uniform time modulation},
  author={Wang, Xuchen and Ptitcyn, Grigorii and Asadchy, VS and D{\'\i}az-Rubio, A and Mirmoosa, Mohammad Sajjad and Fan, Shanhui and Tretyakov, Sergei A},
  journal={Physical Review Letters},
  volume={125},
  number={26},
  pages={266102},
  year={2020},
  publisher={APS}
}

@article{fleury2018non,
  title={Non-reciprocal optical mirrors based on spatio-temporal acousto-optic modulation},
  author={Fleury, Romain and Sounas, DL and Al{\`u}, Andrea},
  journal={Journal of Optics},
  volume={20},
  number={3},
  pages={034007},
  year={2018},
  publisher={IOP Publishing}
}

@article{mirmoosa2023quantum,
  title={Quantum state engineering and photon statistics at electromagnetic time interfaces},
  author={Mirmoosa, MS and Set{\"a}l{\"a}, T and Norrman, A},
  journal={Physical Review Research},
  volume={7},
  number={1},
  pages={013120},
  year={2025},
  publisher={APS}
}

@article{mendoncca2000quantum,
  title={Quantum theory of time refraction},
  author={Mendonca, JT and Guerreiro, A and Martins, Ana M},
  journal={Physical Review A},
  volume={62},
  number={3},
  pages={033805},
  year={2000},
  publisher={APS}
}

@article{morgenthaler1958velocity,
  title={Velocity modulation of electromagnetic waves},
  author={Morgenthaler, Frederic R},
  journal={IRE Transactions on Microwave Theory and Techniques},
  volume={6},
  number={2},
  pages={167--172},
  year={1958},
  publisher={IEEE}
}

@article{Tretyakov2024Bian,
  title = {Time interfaces in bianisotropic media},
  author = {Mirmoosa, M. S. and Mostafa, M. H. and Norrman, A. and Tretyakov, S. A.},
  journal = {Phys. Rev. Res.},
  volume = {6},
  issue = {1},
  pages = {013334},
  number = {13},
  year = {2024},
  month = {Mar},
  publisher = {American Physical Society} 
}

@article{Alu_TL,
  title={Observation of temporal reflection and broadband frequency translation at photonic time interfaces},
  author={Moussa, Hady  and Xu, Gengyu and Yin, Shixiong  and Galiffi, Emanuele  and  Ra’di, Younes and Alù, Andrea},
  journal={Nature Physics},
  volume={19},
  pages={863--868},
  year={2023},
  doi = {10.1038/s41567-023-01975-y}
}

@article{lustig2023timeEx,
  title={Time-refraction optics with single cycle modulation},
  author={Lustig, Eran and Segal, Ohad and Saha, Soham and Bordo, Eliyahu and Chowdhury, Sarah N and Sharabi, Yonatan and Fleischer, Avner and Boltasseva, Alexandra and Cohen, Oren and Shalaev, Vladimir M and others},
  journal={Nanophotonics},
  volume={12},
  number={12},
  pages={2221--2230},
  year={2023},
  publisher={De Gruyter}
}

@article{mendoncca2002time,
  title={Time refraction and time reflection: two basic concepts},
  author={Mendon{\c{c}}a, JT and Shukla, PK},
  journal={Physica Scripta},
  volume={65},
  number={2},
  pages={160},
  year={2002},
  publisher={IOP Publishing}
}

@article{mendoncca2003temporal,
  title={Temporal beam splitter and temporal interference},
  author={Mendon{\c{c}}a, JT and Martins, AM and Guerreiro, A},
  journal={Physical Review A},
  volume={68},
  number={4},
  pages={043801},
  year={2003},
  publisher={APS}
}

@article{vazquez2022shaping,
  title={Shaping the quantum vacuum with anisotropic temporal boundaries},
  author={V{\'a}zquez-Lozano, J Enrique and Liberal, I{\~n}igo},
  journal={Nanophotonics},
  volume={12},
  number={3},
  pages={539--548},
  year={2022},
  publisher={De Gruyter}
}

@article{liberal2023quantum,
  title={Quantum antireflection temporal coatings: quantum state frequency shifting and inhibited thermal noise amplification},
  author={Liberal, I{\~n}igo and V{\'a}zquez-Lozano, J Enrique and Pacheco-Pe{\~n}a, Victor},
  journal={Laser \& Photonics Reviews},
  volume={17},
  number={9},
  pages={2200720},
  year={2023},
  publisher={Wiley Online Library}
}

@article{Agrawal2014RTC,
  title={Reflection and transmission of electromagnetic waves at a temporal boundary},
  author={Xiao, Yuzhe and Maywar, Drew N and Agrawal, Govind P},
  journal={Optics letters},
  volume={39},
  number={3},
  pages={574--577},
  year={2014},
  publisher={Optica Publishing Group}
}

@article{ptitcyn2023time,
  title={Time-modulated circuits and metasurfaces for emulating arbitrary transfer functions},
  author={Ptitcyn, GA and Mirmoosa, Mohammad Sajjad and Hrabar, Silvio and Tretyakov, Sergey A},
  journal={Physical review applied},
  volume={20},
  number={1},
  pages={014041},
  year={2023},
  publisher={APS}
}

@book{PABook, 
author={Gonorovsky, I. S.}, 
title={Radio Circuits and Signals}, 
year={1981}, 
publisher={Mir Publishers}, 
address={Moscow, USSR} 
}

@article{cullen1958travelling, 
title={A travelling-wave parametric amplifier}, 
author={Cullen, AL}, 
journal={Nature}, 
volume={181}, 
number={4605}, 
pages={332--332}, 
year={1958}, 
publisher={Nature Publishing Group UK London} 
}

@article{magierowski2010rf, 
title={RF CMOS parametric downconverters}, 
author={Magierowski, Sebastian and Bousquet, Jean-Francois and Zhao, Zhixing and Zourntos, Takis}, 
journal={IEEE Transactions on Microwave Theory and Techniques}, 
volume={58}, 
number={3}, 
pages={518--528}, 
year={2010}, 
publisher={IEEE} 
}

@article{hedayati2021parametric, 
title={Parametric downconverter for mixer-first receiver front ends}, 
author={Hedayati, Maziar and Yeung, Lap K and Panahi, Mohammadali and Zou, Xiating and Wang, Yuanxun Ethan}, 
journal={IEEE Transactions on Microwave Theory and Techniques}, 
volume={69}, 
number={5}, 
pages={2712--2721}, 
year={2021}, 
publisher={IEEE} 
}

@article{gray2010analytical, 
title={Analytical modeling of microwave parametric upconverters}, 
author={Gray, Blake and Melville, Bob and Kenney, J Stevenson}, 
journal={IEEE Transactions on Microwave Theory and Techniques}, 
volume={58}, 
number={8}, 
pages={2118--2124}, 
year={2010}, 
publisher={IEEE} 
}

@article{Afshari2010noise, 
title={Low-noise parametric resonant amplifier}, 
author={Lee, Wooram and Afshari, Ehsan}, 
journal={IEEE Transactions on Circuits and Systems I: Regular Papers}, 
volume={58}, 
number={3}, 
pages={479--492}, 
year={2010}, 
publisher={IEEE} 
}

@article{qin2014nonreciprocal, 
title={Nonreciprocal components with distributedly modulated capacitors}, 
author={Qin, Shihan and Xu, Qiang and Wang, Yuanxun Ethan}, 
journal={IEEE Transactions on Microwave Theory and Techniques}, 
volume={62}, 
number={10}, 
pages={2260--2272}, 
year={2014}, 
publisher={IEEE} 
}

@article{mirmoosa2019time, 
title={Time-varying reactive elements for extreme accumulation of electromagnetic energy}, 
author={Mirmoosa, Mohammad Sajjad and Ptitcyn, GA and Asadchy, Viktar S and Tretyakov, Sergei A}, 
journal={Physical Review Applied}, 
volume={11}, 
number={1}, 
pages={014024}, 
year={2019}, 
publisher={APS} 
}

@article{PhysRevB.96.155409,
  title = {Optical isolation based on space-time engineered asymmetric photonic band gaps},
  author = {Chamanara, Nima and Taravati, Sajjad and Deck-L\'eger, Zo\'e-Lise and Caloz, Christophe},
  journal = {Phys. Rev. B},
  volume = {96},
  issue = {15},
  pages = {155409},
  numpages = {12},
  year = {2017},
  month = {Oct},
  publisher = {American Physical Society},
  doi = {10.1103/PhysRevB.96.155409},
  url = {https://link.aps.org/doi/10.1103/PhysRevB.96.155409}
}

@article{PhysRevA.79.053821,
  title = {Reflection and transmission of a wave incident on a slab with a time-periodic dielectric function $\epsilon(t)$},
  author = {Zurita-S\'anchez, Jorge R. and Halevi, P. and Cervantes-Gonz\'alez, Juan C.},
  journal = {Phys. Rev. A},
  volume = {79},
  issue = {5},
  pages = {053821},
  numpages = {13},
  year = {2009},
  month = {May},
  publisher = {American Physical Society},
  doi = {10.1103/PhysRevA.79.053821},
  url = {https://link.aps.org/doi/10.1103/PhysRevA.79.053821}
}

@article{caloz2019spacetimeI1,
  title={Spacetime metamaterials—part I: general concepts},
  author={Caloz, Christophe and Deck-L{\'e}ger, Zo{\'e}-Lise},
  journal={IEEE Transactions on Antennas and Propagation},
  volume={68},
  number={3},
  pages={1569--1582},
  year={2019},
  publisher={IEEE}
}

@article{caloz2019spacetimeII2,
  title={Spacetime metamaterials—Part II: Theory and applications},
  author={Caloz, Christophe and Deck-Leger, Zoe-Lise},
  journal={IEEE Transactions on Antennas and Propagation},
  volume={68},
  number={3},
  pages={1583--1598},
  year={2019},
  publisher={IEEE}
}

@article{Shalaev2025spatio, 
title={Spatio-spectral optical fission in time-varying subwavelength layers}, 
author={Jaffray, Wallace and Stengel, Sven and Biancalana, Fabio and Fruhling, Colton Bradley and Ozlu, Mustafa and Scalora, Michael and Boltasseva, Alexandra and Shalaev, Vladimir M and Ferrera, Marcello}, 
journal={Nature Photonics}, 
pages={1--9}, 
year={2025}, 
publisher={Nature Publishing Group UK London} 
}
\end{document}